%% file: article.tex
\documentclass[preprint]{elsarticle}
\usepackage{natbib}
\usepackage{lineno,hyperref}
\usepackage{hypernat}
\usepackage[T1]{fontenc}
\usepackage{graphicx,wrapfig}
\usepackage{subfiles}
\usepackage{textcomp,gensymb}
\usepackage{color}
\usepackage{caption}
\usepackage{subcaption}
\usepackage{amsmath,amssymb,eucal}
\usepackage{amsfonts}
\usepackage{upgreek}
\usepackage{mathrsfs}
\usepackage{stmaryrd}
\usepackage{float}
\usepackage{amsthm}
\usepackage{bm} 
\usepackage{colortbl}
\usepackage{algpseudocode,algorithm}
\usepackage{setspace}
\usepackage{lipsum}
\usepackage[normalem]{ulem}
\usepackage[symbol]{footmisc} 
\floatstyle{plaintop}
\newfloat{mybox}{tbhp}{}
\floatname{mybox}{\bf{Box}}
\theoremstyle{remark}

\let\oldequation\equation
\let\oldendequation\endequation
\renewenvironment{equation}
  {\linenomathNonumbers\oldequation}
  {\oldendequation\endlinenomath}

\newcommand\norm[1]{\left\lVert#1\right\rVert} 

\newcommand{\brm}[1]{{\mathbf #1}}
\newcommand{\bfs}[1]{\mbox{\boldmath{$ #1 $}}}

\newcommand {\gfrac}[2]  {\genfrac{}{}{}{1}{#1}{#2}}
\newcommand\RD{{\rm D}}

\def\Int{\int\limits}

\DeclareMathOperator{\tr}{tr}

\journal{Elsevier}
\begin{document}
\begin{frontmatter}
\title{Lamination-based efficient treatment of weak discontinuities for non-conforming finite element meshes$^\dagger$}
\author[IPPT]{J\k{e}drzej Dobrza\'{n}ski}
\ead{jdobrz@ippt.pan.pl}
\author[IPPT]{Kajetan Wojtacki}
\ead{kajetan.wojtacki@gmail.com}
\author[IPPT]{Stanis{\l}aw Stupkiewicz\corref{cor1}}
\ead{sstupkie@ippt.pan.pl}

\cortext[cor1]{Corresponding author}

\address[IPPT]{Institute of Fundamental Technological Research (IPPT), Polish Academy of Sciences,\\
Pawi\'nskiego 5B, 02-106 Warsaw, Poland.}

\begin{abstract}
When modelling discontinuities (interfaces) using the finite element method, the standard approach is to use a conforming finite-element mesh in which the mesh matches the interfaces. However, this approach can prove cumbersome if the geometry is complex, in particular in 3D. 
In this work, we develop an efficient technique for a non-conforming finite-element treatment of weak discontinuities by using laminated microstructures. The approach is inspired by the so-called composite voxel technique that has been developed for FFT-based spectral solvers in computational homogenization. The idea behind the method is rather simple. Each finite element that is cut by an interface is treated as a simple laminate with the volume fraction of the phases and the lamination orientation determined in terms of the actual geometrical arrangement of the interface within the element. The approach is illustrated by several computational examples relevant to the micromechanics of heterogeneous materials. Elastic and elastic-plastic materials at small and finite strain are considered in the examples. The performance of the proposed method is compared to two alternative, simple methods showing that the new approach is in most cases superior to them while maintaining the simplicity. 
\footnotetext[2]{Published in \emph{Computers \& Structures}, \textbf{291}, 107209, 2024, doi: 10.1016/j.compstruc.2023.107209}
\end{abstract}
\begin{keyword}
finite element method \sep interface \sep weak discontinuity \sep laminate \sep homogenization \sep elasticity \sep plasticity
\end{keyword}

\end{frontmatter}


\section{Introduction}\label{intro}

\subfile{sections/01_Introduction}

%
%
\section{Preliminaries}\label{theory}
\subfile{sections/02_Theory}

%
%
\section{Laminated element technique}\label{LEM}
\subfile{sections/03_LEM}

%
%
\section{Illustrative examples}\label{numresults}

\begin{figure}[h]
\centerline{\includegraphics[scale=0.33]{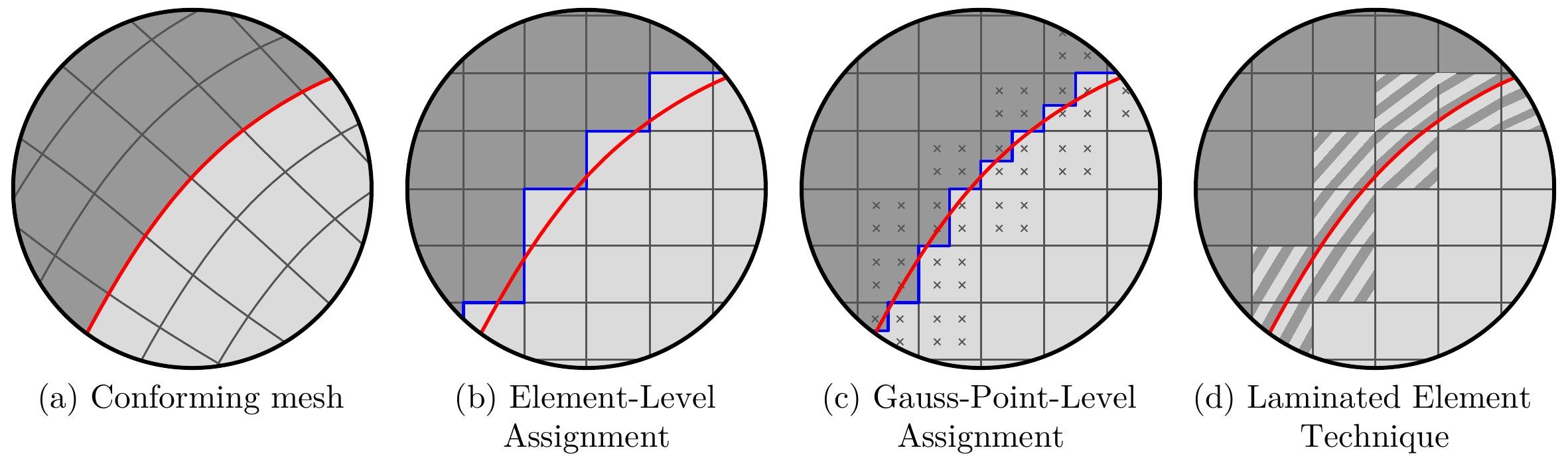}}
\caption{Discretization approaches employed in this work: (a) conforming mesh, (b) element-level assignement (ELA), (c) Gauss-point-level assignement (GPLA), and (d) laminated element technique (LET). In ELA (resp.\ GPLA), the whole element (resp.\ Gauss point) belongs to a single phase that is determined by the value of the level-set function in the element centre (resp.\ at the Gauss point). 
}
\label{scheme}
\end{figure}

Performance of LET is examined in this section through several numerical examples. In all examples, we use a regular quadrilateral (2D) or hexahedral (3D) mesh. 
In all cases, the results are compared to those obtained using two simple non-conforming mesh approaches that {will be} referred to as ELA (element-level assignement) and GPLA (Gauss-point-level assignement), see Fig.~\ref{scheme}. 
Whenever applicable a conforming mesh is also used.

%
%
%
%

{As illustrated in Fig.~\ref{scheme}(b), in ELA, the whole element is assigned to one of the phases, and this approach, sometimes called digital-image-based FEM or voxel-based FEM, is commonly used for segmented 2D and 3D images of the microstructure or for rasterized representation of the microstructure, e.g.\ \citep{Keyak1990,Terada1997,Lian2012}. 
In GPLA, Fig.~\ref{scheme}(c), individual integration (Gauss) points are assigned to one of the phases according to the position of the integration point, possibly combined with an increased number of integration points, e.g.\ \citep{Moes2003,Essongue2020}. 
In the context of voids and free boundaries, this latter approach is closely related to the finite cell method \citep{Parvizian2007}.}

Finite-element implementation and computations have been performed using the \emph{AceGen/AceFEM} system \citep{Korelc2009,KorelcWriggers2016}.

\subsection{Elastic inclusion}\label{inclusion}

\subfile{sections/04_Inclusion2DSSE}

%
%
\subsection{Compatible eigenstrain at a planar interface}

\subfile{sections/05_Eigen2DSSE}

%
%
\subsection{Elastic inclusion with varying radius}
\subfile{sections/06_Inclusion3DSSE}

%
%
\subsection{Hyperelastic woven microstructure}\label{sec:woven}
\subfile{sections/07_Wovencell}

%
%
\subsection{Elasto-plastic composite}\label{sec:EPcomp}
\subfile{sections/08_Inclusion2DFSEP}

%
%
\section{Conclusion}
A simple approach has been developed for an improved treatment of weak discontinuities for a non-conforming spatial discretization within the finite element method. The use of a non-conforming finite-element mesh has several advantages over the conforming one, in particular, in 3D and for complex geometries, but comes at the cost of loss of accuracy caused by an inexact representation of the geometry of the phases. The proposed approach is thus aimed at improving the accuracy while maintaining the simplicity such that implementation is carried out solely at the element level and no additional global degrees of freedom are introduced.

In the proposed laminated element technique (LET), each finite element that is cut by an interface is treated as a laminate of the two involved phases with the volume fractions of the phases equal to their volume fractions within the element and with the lamination orientation specified by the orientation of the interface. No treatment is applied to the {remaining} elements, i.e., {those} that are not cut by an interface. The approach is inspired by the composite voxel technique in FFT-based homogenization \citep{Gelebart2015,Kabel2015,Kabel2016,Kabel2017,Mareau2017,Keshav2022}.

The approach is general in the sense that each phase may be governed by an arbitrary material model. The constitutive behaviour of each laminated element results from the closed-form, exact micro-to-macro transition relations for simple laminates. For nonlinear material behaviour (e.g., plasticity, finite deformations), a set of nonlinear equations must be solved at each Gauss point. This makes the effective constitutive model within the laminated elements somewhat more complex, but an efficient computational implementation is possible, including consistent linearization, as illustrated in the case of finite-strain plasticity.

Several numerical examples have been studied and the proposed approach has been shown to be, in most cases, superior in terms of accuracy to two alternative methods in which the whole element or individual Gauss points are assigned to a specific phase. However, the rate of convergence with mesh refinement is not improved, {unlike in} more sophisticated approaches, such as X-FEM. On the other hand, it is an important feature of the proposed approach that the response is a continuous function of the position of the interface, which opens a possibility of its application in various problems involving moving interfaces.

%
%
\paragraph{Acknowledgment}
This work has been partially supported by the National Science Centre (NCN) in Poland through Grant No.\ 2018/29/B/ST8/00729.
%
%
\appendix
\subfile{sections/99_AppendixA}

%
%
\bibliography{bibliografia}
\end{document}

%% file: sections/01_Introduction.tex
Solving partial differential equations on complex geometries plays a dominant role in many problems of interest in computational solid mechanics. Analytical solutions 
that consider possible geometrical heterogeneities are not available in most cases. Computational approaches are thus indispensable, the finite element method (FEM) being the most general, most powerful and most popular computational tool for numerical simulations in various areas of engineering.

Historically, FEM relies on geometry-dependent computational grids (body-fitted or conforming meshing). 
However, due to the complexity of investigated geometries, especially internal heterogeneities, a fine conforming discretisation can be quite hard to use. 
One reason is the limited computational cost that can be afforded. 
Although algorithms for finite element meshing in 2D are quite efficient and well-established, the mesh generation for 3D problems still remains a rather cumbersome and time-consuming task. 
This problem is particularly pronounced when discretising geometries that contain discontinuities or objects (details) with diverse characteristic lengths. 
Secondly, some additional manual input is often needed or use of specialized mesh generation software.
Thirdly, for 3D models, especially large-scale ones, {unstructured meshes, which are necessarily needed to represent complex geometries, may lead to additional difficulties associated with the assembly and solution of the respective finite-element equations.}
And finally, even if the previous obstacles can be somehow overcome, {additional effort may be needed} during post-processing.
Overall, mesh generation may become the most time-consuming process in the preprocessing step of numerical modelling, and dealing with conforming meshes for complex geometries may significantly increase the computational expense.

A possible approach to evade the mesh generation problem is to use a non-conforming mesh. 
In general, the respective methods use a structured mesh or a simple unstructured mesh generated in the domain defined by the external boundary of the modelled geometry. 
Generating such a mesh 
is then a straightforward process. 
However, the internal details of the geometry must somehow be treated, and several approaches have been developed for that purpose, as discussed below. 

The focus of this work is on \emph{weak discontinuities}, i.e., on the situation in which the primal variable (e.g., the displacement field) is continuous at the interface and discontinuous are its derivatives {and related quantities} (e.g., strains and stresses) when the material properties (e.g., elastic moduli) suffer discontinuity at the interface. 
This is in contrast to strong discontinuities, such as cracks, when the primal variable may be discontinuous at the interface (of possibly unknown and evolving shape). 
Actually, several methods have been primarily developed for strong discontinuities and have then been adapted to weak discontinuities.

This is, for instance, the case of the extended finite element method (X-FEM) {initially} developed for modelling crack propagation independent of the underlying finite-element mesh \citep{Belytschko1999,Moes1999}, see also two methods that are closely related to X-FEM, namely CutFEM \citep{Burman2014} and phantom node method (PNM) \citep{Song2006}, and are sometimes considered just versions of X-FEM. 
It has been subsequently shown that X-FEM can be successfully used also for modelling complex internal geometries (weak discontinuities) of the geometry independent of the finite-element mesh \citep{Sukumar2001,Belytschko2002,Moes2003}. 
With increasing popularity of the isogeometric analysis (IGA) \citep{Hughes2005}, the X-FEM approach has been also combined with IGA \citep{DeLuycker2011,Ghorashi2011}, including XIGA for weak discontinuities and multimaterial problems \citep{Tambat2012,Noel2022}.

In X-FEM, the inner surfaces (e.g., material interfaces, cracks) are defined implicitly using level set functions \citep{Osher1988}. 
Enrichment functions are then employed to modify the finite-element approximation of the displacement field such that the discontinuity is represented on a non-matching mesh. 
As a result, the {optimal} convergence rate can be achieved. 
The beneficial features of X-FEM come at the cost that additional global degrees of freedom are introduced (those associated with the enrichment shape functions). 
Moreover, integration must be performed accurately on the elements cut by an interface and, for this purpose, the elements are triangulated such that the subdomains match the interfaces. 
This becomes even more complex when more than one interface passes through an element, which is not so improbable, for instance, in the case of small inclusions or multi-material problems.
Overall, implementation of X-FEM is not straightforward, particularly in 3D, and cannot be performed solely at the element level.

Additional deformation modes, in a sense similar to the enrichment functions of X-FEM, are also introduced in the immersed interface FEM (IIFEM) \citep{Li2005,Lin2012} and in the augmented finite element method (AFEM) \citep{Liu2014,Essongue2020}. 
The difference is that, unlike in X-FEM, the enrichment functions are not continuous at the inter-element boundaries, which implies that the optimal convergence rate cannot be achieved. To improve convergence, the inter-element compatibility is enforced in a version of
IIFEM \citep{Li2003}, which then bears some similarity to X-FEM. 
In the {incompatible} case, the additional degrees of freedom associated with the additional deformation modes can be condensed at the element level, hence no additional global degrees of freedom are introduced.

Common to the approaches discussed above is the ``small cut-cell'' problem that may appear when a small part of the element is cut by an interface. This may lead to a large condition number of the algebraic system to be solved and may deteriorate numerical stability of the resulting algorithms, thus additional stabilization techniques are needed.

A different approach is adopted in the shifted interface method (SIM) \citep{Li2019,LiKangan2021} in which the interface is shifted to a nearby inter-element boundary. At the same time, to compensate {for} the error introduced by shifting the interface, the interface jump (compatibility) conditions are applied at the surrogate interface in a modified form resulting from the Taylor expansion of the original jump conditions. 
{As a result, the optimal convergence rate can be achieved \citep{Li2019}.}

In this work, we develop a simple method for improved treatment of weak discontinuities in non-conforming FEM discretization. 
The inspiration came from recent developments in the FFT-based methods in computational homogenization. 
In this class of approaches \citep{Moulinec1998}, a periodic unit cell is discretized into a regular array of voxels, hence complex-shaped interfaces cannot be represented exactly. 
In order to increase the accuracy of FFT-based homogenization, the idea of \emph{composite voxels} has been introduced in \citep{Brisard2010}, see also \citep{Toulemonde2008} for a related approach in the context of FEM, and was the first attempt to use a homogenization technique 
{to} prescribe effective mechanical properties to the voxel that contains an interface between two phases.
This idea was further developed by considering the composite voxel to be represented by a {\emph{laminated microstructure}} and characterized by the corresponding effective properties \citep{Gelebart2015,Kabel2015}. It has been shown that these laminate voxels significantly improve the accuracy of the method as compared to the composite voxels employing simple Voigt and Reuss bounds \citep{Kabel2015}. Further related developments in the context of FFT-based homogenization include the extension to inelastic problems \citep{Mareau2017,Kabel2017} and to the finite-strain framework \citep{Kabel2016,Keshav2022}.

In this work, the idea of laminate-based composite voxels is applied in the context of the finite element method. In short, each element that is cut by an interface is treated as a laminate composed of the two phases with the volume fraction and lamination orientation determined by the actual geometry of the interface within the element, see Section~\ref{LEM}. The method, which we call the laminated element technique (LET), does not introduce any additional global degrees of freedom and can be implemented solely at the element level. The performance of the method is examined through several numerical experiments, see Section~\ref{numresults}.

{As illustrated by the numerical examples, in terms of accuracy, the proposed method cannot compete, and is not aimed to compete, with more sophisticated methods, such as X-FEM, which can achieve the optimal convergence rate typical for conforming-mesh FEM. However, the beneficial features of LET, as compared X-FEM, are its simplicity and the ease of implementation. At the same time, it delivers an improved accuracy, as compared to two simple non-conforming mesh approaches examined 
as a reference.}

%% file: sections/02_Theory.tex
\subsection{Compatibility conditions at a bonded interface}


Consider a body occupying, in the reference configuration, domain $\Omega$ that is divided into two subdomains $\Omega_1$ and $\Omega_2$ ($\Omega=\Omega_1\cup\Omega_2$, $\Omega_1\cap\Omega_2=\emptyset$) with homogeneous material properties within {subdomains} $\Omega_1$ and $\Omega_2$.
The interface separating $\Omega_1$ and $\Omega_2$ is assumed to be smooth and is denoted by $\Sigma$ with $\bm{N}$ 
denoting the unit normal outward to $\Omega_1$.

Deformation of the body is described by the deformation mapping $\bm{\varphi}$ such that $\bm{x}=\bm{\varphi}(\bm{X})=\bm{X}+\bm{u}(\bm{X})$ where $\bm{X}$ and $\bm{x}$ denote the position of a material point in the reference and current configuration, respectively, and $\bm{u}$ is the displacement field.
The deformation mapping is assumed to be continuously differentiable in $\Omega_1$ and {in} $\Omega_2$ and continuous on $\Sigma$. The deformation gradient $\bm{F}=\nabla\bm{\varphi}$ can thus be defined in $\Omega_1$ and {in} $\Omega_2$, while continuity of $\bm{\varphi}$ implies the following \emph{kinematic compatibility condition} on $\Sigma$, e.g.\ \citep{Silhavy1997},
\begin{equation}
\label{eq:comp:F}
    \bm{F}_2-\bm{F}_1=\bm{c}\otimes\bm{N} ,
\end{equation}
where $\bm{F}_1$ and $\bm{F}_2$ are the two limiting values of the deformation gradient at the interface $\Sigma$, $\bm{c}$ is a vector, $\nabla$ denotes the gradient in the reference configuration, {and $\otimes$ denotes the diadic product}.

The equilibrium equation, $\nabla\cdot\bm{P}^{\rm T}=\bm{0}$ ($P_{ij,j}=0$), is formulated in $\Omega_1$ and {in} $\Omega_2$ in terms of the Piola (first Piola--Kirchhoff) stress tensor $\bm{P}=J\bm{\sigma}\bm{F}^{\rm -T}$, where $\bm{\sigma}$ is the Cauchy stress tensor, and $J=\det\bm{F}$. At the interface, equilibrium requires that the traction vector is continuous,
\begin{equation}
\label{eq:comp:P}
  (\bm{P}_2-\bm{P}_1)\bm{N}=\bm{0} ,
\end{equation}
which can be written also in the current configuration, $(\bm{\sigma}_2-\bm{\sigma}_1)\bm{n}=\bm{0}$, where $\bm{n}$ is the normal to the interface in the current configuration. 
The interface $\Sigma$ is thus a surface of weak discontinuity at which the displacement is continuous, but the deformation gradient $\bm{F}$ and the Piola stress $\bm{P}$ may suffer discontinuity due to the jump in the material properties at $\Sigma$.

In the {small-strain (geometrically linear)} framework, the compatibility conditions, Eqs.~\eqref{eq:comp:F} and~\eqref{eq:comp:P}, take the following form,
\begin{equation}
\label{eq:comp:SS}
\bm{\varepsilon}_2-\bm{\varepsilon}_1=\frac12(\bm{c}\otimes\bm{n}+\bm{n}\otimes\bm{c}) , \qquad
(\bm{\sigma}_2-\bm{\sigma}_1)\bm{n}=\bm{0} ,
\end{equation}
where $\bm{\varepsilon}$ is the usual infinitesimal strain tensor (the symmetric part of the displacement gradient), and no distinction is made between the current and reference configurations, hence $\bm{n}=\bm{N}$.

\subsection{Simple laminate}
\label{sec:simple}

{A simple} laminate is a microstructure composed of layers of two phases (materials) separated by parallel planar interfaces. Under the usual assumption of separation of scales, strains and stresses are homogeneous within each individual layer and are identical in all layers of the same phase. The microstructure is then fully characterized by the volume fractions of the phases, $\eta_1=1-\eta$ and $\eta_2=\eta$, {where} $0\leq\eta\leq1$, and by the interface normal $\bm{N}$, all referred to the reference configuration.

Since the strains and stresses are piecewise homogeneous, the macroscopic 
deformation gradient $\bar{\bm{F}}=\left\{ \bm{F} \right\}$ and {the macroscopic} Piola stress $\bar{\bm{P}}=\left\{ \bm{P} \right\}$ are obtained as simple weighted averages of the respective local quantities,
\begin{equation}
\label{eq:macro:PF}
\bar{\bm{F}}={(1-\eta)\bm{F}_1+\eta\bm{F}_2} , \qquad
\bar{\bm{P}}={(1-\eta)\bm{P}_1+\eta\bm{P}_2} ,
\end{equation}
where $\left\{ \boxempty \right\}$ denotes the average over the representative volume element in the reference configuration. Here, $\bm{F}_i$ and $\bm{P}_i$ denote the local quantities within the individual phases.

The compatibility conditions, Eqs.~\eqref{eq:comp:F} and~\eqref{eq:comp:P}, and the averaging rules, Eq.~\eqref{eq:macro:PF}, complemented by the local constitutive laws of the phases, are sufficient to determine the macroscopic constitutive law relating the macroscopic quantities, $\bar{\bm{F}}$ and $\bar{\bm{P}}$. This is illustrated below for the case of hyperelastic constituents. The general case of elastic-plastic phases is discussed in~\ref{app:ep}, where the corresponding computational scheme is presented including the structure of the nested iterative-subiterative scheme and its linearization.

A hyperelastic material model is fully defined by specifying the elastic strain energy function. Denoting by $W_i=W_i(\bm{F}_i)$ the elastic strain energy of phase $i$, the corresponding Piola stress $\bm{P}_i$ is obtained as
\begin{equation}
\bm{P}_i = \frac{\partial W_i(\bm{F}_i)}{\partial\bm{F}_i} .
\end{equation}
While the requirement of objectivity implies that $W_i$ is in fact a function of the right Cauchy--Green tensor $\bm{C}_i=\bm{F}_i^{\rm T}\bm{F}_i$, it is convenient here to keep the deformation gradient $\bm{F}_i$ as the argument of $W_i$. Since the constituent phases are hyperelastic, the laminate is also a hyperelastic material. Its behaviour is thus governed by the corresponding (macroscopic) elastic strain energy function that depends on the macroscopic deformation gradient $\bar{\bm{F}}$, as discussed below. 

Using the kinematic compatibility condition~\eqref{eq:comp:F} and the averaging rule~\eqref{eq:macro:PF}${}_1$, the local deformation gradients $\bm{F}_i$ can be expressed in terms of the macroscopic deformation gradient $\bar{\bm{F}}$ and (yet unknown) vector $\bm{c}$,
\begin{equation}
\label{eq:Fi}
\bm{F}_1={\bar{\bm{F}}-\eta\bm{c}\otimes\bm{N}} , \qquad
\bm{F}_2={\bar{\bm{F}}+(1-\eta)\bm{c}\otimes\bm{N}} .
\end{equation}
The macroscopic elastic strain energy $\bar{W}=\left\{  W \right\}$ is thus also a function of $\bar{\bm{F}}$ and $\bm{c}$,
\begin{equation}
\bar{W}(\bar{\bm{F}},\bm{c})={(1-\eta) W_1(\bm{F}_1)+\eta W_2(\bm{F}_2)} .
\end{equation}

The unknown vector $\bm{c}$ can now be determined from the compatibility condition~\eqref{eq:comp:P}. 
The local deformation gradients $\bm{F}_i$ specified by Eq.~\eqref{eq:Fi} are kinematically admissible since this representation satisfies the kinematic compatibility condition~\eqref{eq:comp:F} and the averaging rule~\eqref{eq:macro:PF}${}_1$ by construction. Accordingly, the local equilibrium of the laminate, expressed by the compatibility condition~\eqref{eq:comp:P}, corresponds to the minimum of the macroscopic elastic strain energy with respect to $\bm{c}$ (at prescribed $\bar{\bm{F}}$). Indeed, the condition of stationarity of $\bar{W}$ gives
\begin{equation}
\label{eq:min:c}
\bm{0}=\frac{\partial\bar{W}}{\partial\bm{c}}
  ={(1-\eta)\frac{\partial W_1}{\partial\bm{F}_1}\frac{\partial\bm{F}_1}{\partial\bm{c}}
    +\eta\frac{\partial W_2}{\partial\bm{F}_2}\frac{\partial\bm{F}_2}{\partial\bm{c}}
  =\eta(1-\eta)(\bm{P}_2-\bm{P}_1)\bm{N}} ,
\end{equation}
which is equivalent to Eq.~\eqref{eq:comp:P} in the non-trivial case of $0<\eta<1$. 

Eq.~\eqref{eq:comp:P} is a nonlinear equation to be solved for the unknown vector $\bm{c}$, 
for instance, using the Newton method. 
The solution of Eq.~\eqref{eq:comp:P} depends (implicitly) on $\bar{\bm{F}}$ so that we have $\bm{c}=\bm{c}(\bar{\bm{F}})$. The macroscopic elastic strain energy can thus be written as a function of $\bar{\bm{F}}$ only,
\begin{equation}
\bar{W}^\ast(\bar{\bm{F}})=\bar{W}(\bar{\bm{F}},\bm{c}(\bar{\bm{F}})) ,
\end{equation}
such that $\bar{W}^\ast$ indeed governs the macroscopic response of the laminate,
\begin{equation}
\label{eq:Pbar}
\bar{\bm{P}}=\frac{\partial\bar{W}^\ast}{\partial\bar{\bm{F}}} .
\end{equation}
To prove Eq.~\eqref{eq:Pbar}, we observe that
\begin{equation}
\frac{\partial\bar{W}^\ast}{\partial\bar{\bm{F}}}
  =\frac{\partial\bar{W}}{\partial\bar{\bm{F}}}+\frac{\partial\bar{W}}{\partial\bm{c}}\frac{\partial\bm{c}}{\partial\bar{\bm{F}}}
  ={(1-\eta)\frac{\partial W_1}{\partial\bm{F}_1}\frac{\partial\bm{F}_1}{\partial\bar{\bm{F}}}
    +\eta\frac{\partial W_2}{\partial\bm{F}_2}\frac{\partial\bm{F}_2}{\partial\bar{\bm{F}}}
  =(1-\eta)\bm{P}_1+\eta\bm{P}_2
  =\bar{\bm{P}}} ,
\end{equation}
where $\partial\bar{W}/\partial\bm{c}=\bm{0}$ in view of Eq.~\eqref{eq:min:c}. Computation of the tangent moduli tensor is not discussed here since it is discussed in \ref{app:ep} in a more general setting.

In the case of linear elastic phases, the laminate is also a linear elastic material fully characterized by a fourth-order tensor of overall elastic moduli. Closed-form formulae for the overall moduli can be found
in \citep{Stupkiewicz2007}.

%% file: sections/03_LEM.tex
\begin{figure}[h]
\centerline{\includegraphics[scale=0.18]{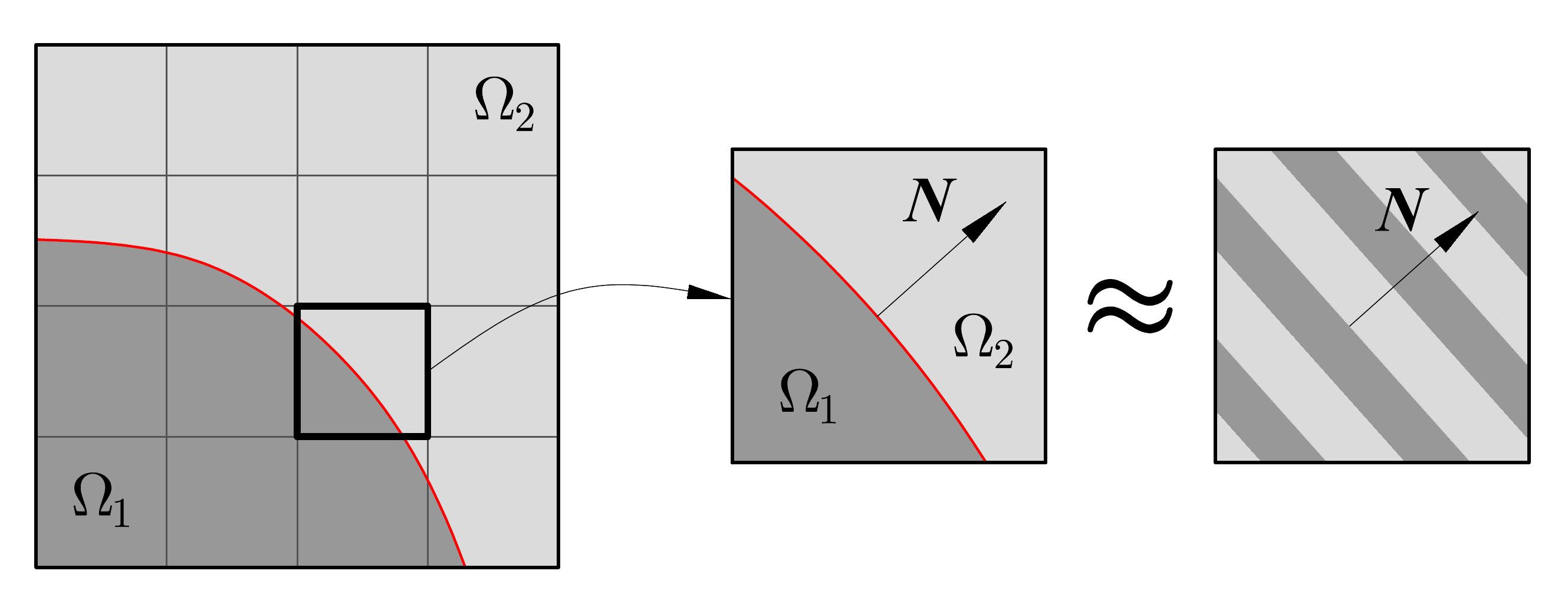}}
\caption{Laminated element technique (LET): the element that is cut by an interface is treated as a simple laminate with the volume fraction and lamination orientation specified by the actual geometry of the interface within the element. {$\bm{N}$ is the unit normal to the interface.}
}
\label{scheme:LET}
\end{figure}

%
%
The main idea of the laminated element technique (LET) is rather simple: the finite element that is cut by {an} interface is treated as a simple laminate composed of the two phases involved, see Fig.~\ref{scheme:LET}. {No} treatment is applied to the elements that fully belong to one phase. 
In each laminated element, the volume fraction of the phases and the lamination orientation are determined according to the actual geometry of the interface within the element, as described in detail below. 

The geometry is defined by the level-set function $\phi$ defined over the whole domain $\Omega$,
\begin{equation}
\phi: \; \Omega \subset \mathbb{R}^n \rightarrow \mathbb{R},
\end{equation}
such that $\phi<0$ corresponds to phase 1, $\phi>0$ corresponds to phase 2, and the {interface $\Sigma$} separating the two phases is represented by the zero level set,
\begin{equation}
{\Sigma}=\left\{ \bm{X} \in \Omega \subset \mathbb{R}^n \, \mid \, \phi \left(\bm{X}\right)=0 \right\} ,
\end{equation}
where $n =2,3$ is the space dimension. 
It is desirable that the level-set function $\phi$ be smooth and (approximately) proportional to the signed distance from the interface in the nearest neighbourhood of interface (within the range of one finite element) so that the interface is correctly approximated by the zero level set of the finite-element approximation $\phi^h$ of the level-set function $\phi$,
\begin{equation}
\phi^h=\sum_i N_i \phi_i ,
\end{equation}
where $N_i$ are the usual finite-element basis functions and $\phi_i$ are the nodal values.

As mentioned in the previous section, a simple laminate is uniquely defined by two quantities, namely the volume fraction $\eta$ ({$\eta_1=1-\eta$, $\eta_2=\eta$}) and the unit vector $\bm{N}$ normal to the interfaces separating the phases (in the reference configuration). 
These two quantities are determined locally within each laminated element in terms of the level-set function $\phi^h$ and are assumed constant within each element.

There is some ambiguity concerning determination of the volume fraction. Exact integration of the volume (or area in 2D) is not possible in the general case and is not needed considering the approximation introduced {by LET anyway}.  
In this work, {we only consider four-node quadrilateral elements in 2D and eight-node hexahedral elements in 3D, and} we use the following formula for the volume fraction $\eta^e=\eta^e_{2}$ in the $e$-th element,
\begin{equation}\label{vfrac}
\eta^e=\frac{\sum_{i=1}^{n_\mathrm{n}} \langle \phi_i^e\rangle}{\sum_{i=1}^{n_\mathrm{n}} \left\lvert \phi_i^e \right\rvert},
\end{equation}
where $\phi_i^e$ are the nodal values of the level-set function {$\phi$} in the element, $n_\mathrm{n}$ is the number of nodes in the element, and $\langle \boxempty \rangle = \frac{1}{2}\left( \boxempty + \left\lvert \boxempty \right\rvert \right) $ denotes the Macaulay brackets.

{In 2D, if the considered element is a rectangle and the interface is a straight line cutting two opposite edges of the rectangle, then formula~\eqref{vfrac} gives an exact value of the volume fraction.
Likewise, in 3D, formula~\eqref{vfrac} is exact for a planar interface cutting four parallel edges of an element of the shape of a rectangular cuboid. 
Otherwise, in particular, when the edges are cut differently, formula \eqref{vfrac} is approximate. 
For a non-planar interface, the error decreases with mesh refinement since then the interface effectively tends to be more planar. 
We have made some efforts to generalize formula~\eqref{vfrac} to improve its accuracy for elements of arbitrary, non-rectangular shape, for instance, by including the usual quadrature weights at the element nodes, but the simple formula~\eqref{vfrac} has been found more accurate. 
Note that, for a planar interface in 2D, the volume fraction can be computed in closed form using the general formula for the area of a polygon in terms of the coordinates of the vertices. However, this formula does not generalize to the 3D case. 
The general and simple formula~\eqref{vfrac} is thus adopted in this work, while clearly this part of the model can be replaced by another suitable formulation.}

The unit normal vector {$\bm{N}^e$} is calculated as the normalized gradient of the level-set function $\phi^h$ at the element centre $\bm{X}_0^e$,
\begin{equation}\label{normal}
\bm{N}^e=\frac{{\nabla} {\phi}^h\left( \bm{X}_{0}^e \right)}{\| {\nabla} {\phi}^h\left( \bm{X}_{0}^e \right) \|} .
\end{equation}

Knowing the volume fraction and the interface normal, the overall constitutive response of the laminate can be readily obtained by applying the micro-to-macro transition, as described in Section~\ref{sec:simple}.

The above construction, Eqs.~\eqref{vfrac} and \eqref{normal}, is applicable when the element is cut by one interface only. 
The approach can be generalized to the case of more interfaces (and more phases) by introducing additional level-set functions, each corresponding to one interface. 
A higher-rank laminate can then be considered with the micro-macro transition applied in a hierarchical manner or, if there are only two phases, a simple laminate can be considered with the volume fraction {equal to the total volume fraction of the phases within the element} and {with the} lamination orientation obtained by averaging those corresponding to each interface. The latter approach is employed in the example considered in Section~\ref{sec:woven}.

%% file: sections/04_Inclusion2DSSE.tex
\begin{figure}[H]
    \centerline{
    \begin{tabular}{ccc} 
        \includegraphics[width=0.38\textwidth]{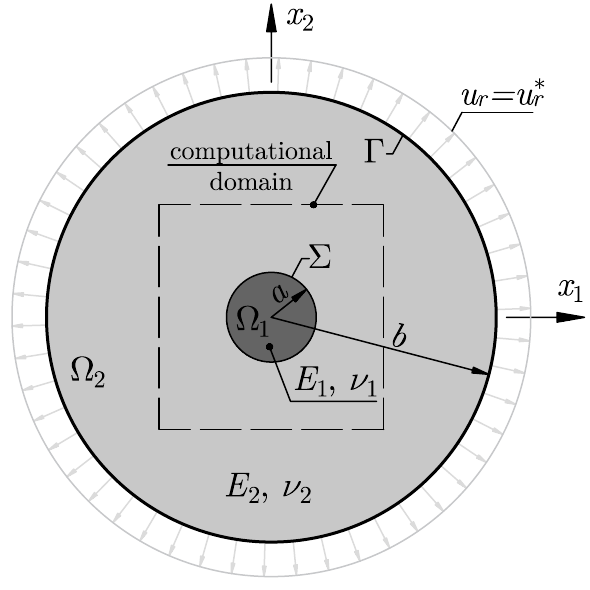} & &
        \includegraphics[width=0.35\textwidth]{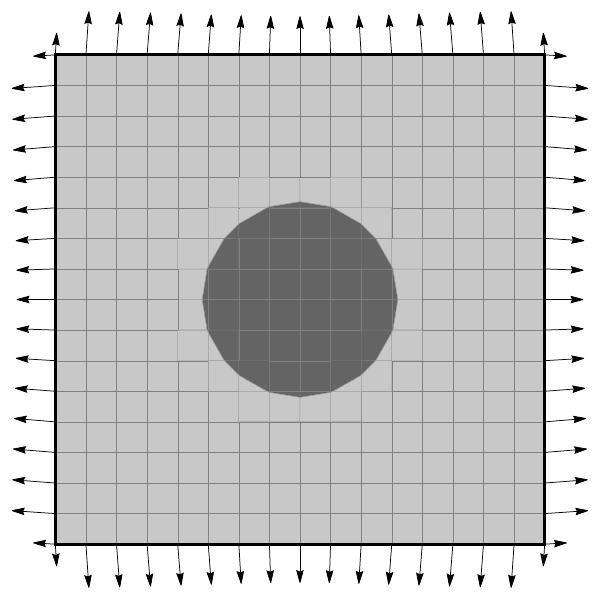} \\
        {\footnotesize (a)}~~ & ~~ & {\footnotesize (b)}
    \end{tabular}
    }
    \caption{Elastic inclusion problem: (a) scheme of the problem; (b) computational domain with a regular (non-conforming) mesh of quadrilateral elements ($16\times16$ elements, $h=0.125$). 
    The interface $\Sigma$ is approximated by the zero level set, $\phi^h=0$. The arrows in panel (b) represent the nodal forces applied to the boundary nodes, which are calculated from the traction resulting from the analytical solution.}
    \label{inclusionscheme}
\end{figure}

In this section, a 2D elastic inclusion problem in the small-strain framework is considered.
The problem is adopted from {\citep{Sukumar2001,Moes2003}}. 
Fig.~\ref{inclusionscheme}(a) shows a body that consists of two domains, $\Omega_1$ (inclusion) and $\Omega_2$, with the elastic constants ($E_1$, $\nu_1$) and ($E_2$, $\nu_2$) that are constant within each domain and suffer discontinuity at the (bonded) {interface $\Sigma$}. 
The radius of the inclusion and the outer radius are denoted by $a$ and $b$, respectively. 
The loading is applied by prescribing the radial displacement {$u_r=u_r^\ast$} 
and zero circumferential displacement $u_{\theta}=0$ at the outer {boundary $\Gamma$}. 
This problem admits an analytical solution that can be found in \citep{Sukumar2001}.  

%
%
%
%

In the computational model, a square-shaped domain with the inclusion in the centre is considered, which is discretized into a regular mesh of isoparametric four-node elements. 
The dimensions of the computational domain are $L\times L$, where $L=2$, while {the parameters specifying the reference problem are adopted as $a=0.4$ and $b=2$.} 
To ensure equivalence with the model described above, the traction resulting from the analytical solution is applied on the boundary of the computational domain. 
Additionally, appropriate displacement boundary conditions are imposed to prevent rigid-body motion. 

\begin{figure}[H]
\centerline{
  \begin{tabular}{cc}
    \includegraphics[width=0.4\textwidth]{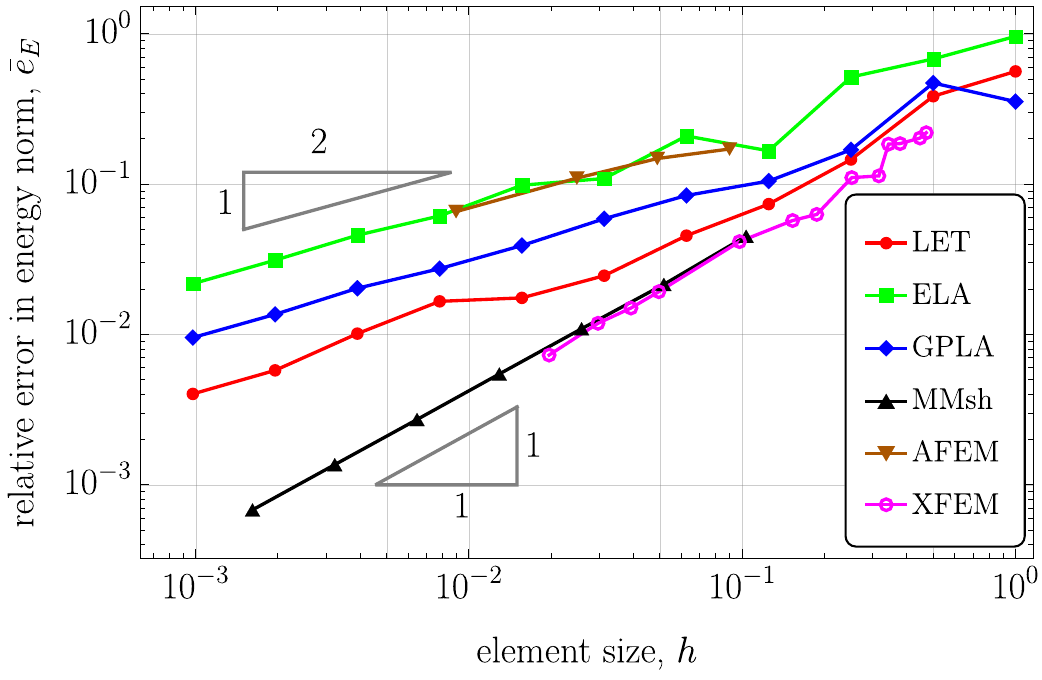} &
    \includegraphics[width=0.4\textwidth]{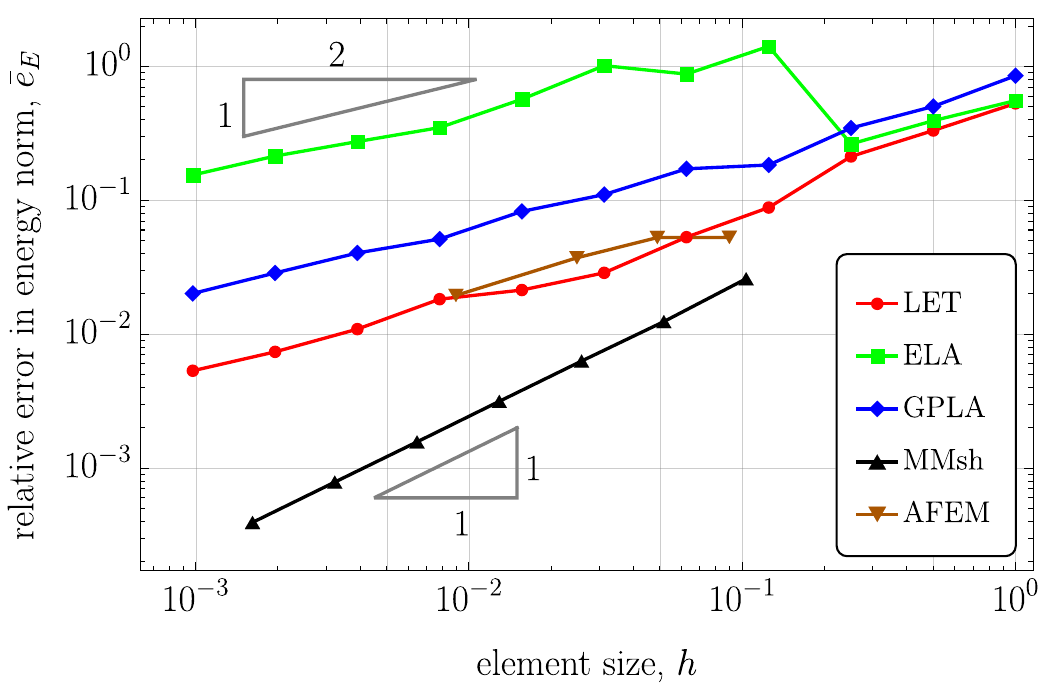} \\
    \hspace*{3em}{\footnotesize (a)} & \hspace*{3em}{\footnotesize (b)}
  \end{tabular}
  }
\caption{{Elastic inclusion problem: rate of convergence in energy norm for (a) soft inclusion ($E_2/E_1=10$) and (b)~hard inclusion ($E_2/E_1=0.005$). The results obtained for a matching mesh of four-node quadrilateral elements are labelled `MMsh'. The results for X-FEM are taken from \citep{Moes2003}, and those for A-FEM from \citep{Essongue2020}.}
}\label{inclCR}
\end{figure}

To examine the performance of LET, the problem is solved for several mesh densitites with the element size $h$ varying between $h=1$ (very coarse mesh with $2\times2$ elements) and $h\approx0.001$. 
{The} elastic constants are adopted as $E_1=1$, $\nu_1=0.25$, $E_2=10$, and $\nu_2=0.3$ {(`soft inclusion' case) and as $E_1=200$, $\nu_1=0.25$, $E_2=1$, and $\nu_2=0.3$ (`hard inclusion' case). 
Convergence of the error is shown in Fig.~\ref{inclCR}. The error is here} defined as the relative error in energy norm, {as in \citep{Moes2003,Essongue2020},}
\begin{equation}\label{relerror}
{\bar{e}_{E}
=\frac{1}{\left(\int_{\Omega} 2 W(\bm{\varepsilon}^{\mathrm{exact}})\,\mathrm{d}\Omega \right)^{1/2}} \norm{\bm{u}^h-\bm{u}^{\text{exact}}}_{E(\Omega)}
=\left(\frac{\int_{\Omega} W(\bm{\varepsilon}^h-\bm{\varepsilon}^{\mathrm{exact}})\,\mathrm{d}\Omega}{\int_{\Omega} W(\bm{\varepsilon}^{\mathrm{exact}})\,\mathrm{d}\Omega}\right)^{1/2},}
\end{equation}
where $W$ is the elastic strain energy density function, {$\bm{u}^{\text{exact}}$ and $\bm{\varepsilon}^{\text{exact}}$ are the exact displacement and strain} obtained from the analytical solution, and {$\bm{u}^h$ and $\bm{\varepsilon}^h$ are the displacement and strain} resulting from the computational model. 
The integrals are evaluated by applying the $2\times2$ Gauss quadrature. 
Note that, in the case of LET, the local strains in each phase are known at the Gauss points of the laminated elements, and the respective local strains are used to evaluate the error. 

%
%
%
%

\begin{figure}[H]
\centerline{
  \begin{tabular}{cc}
    \includegraphics[width=0.4\textwidth]{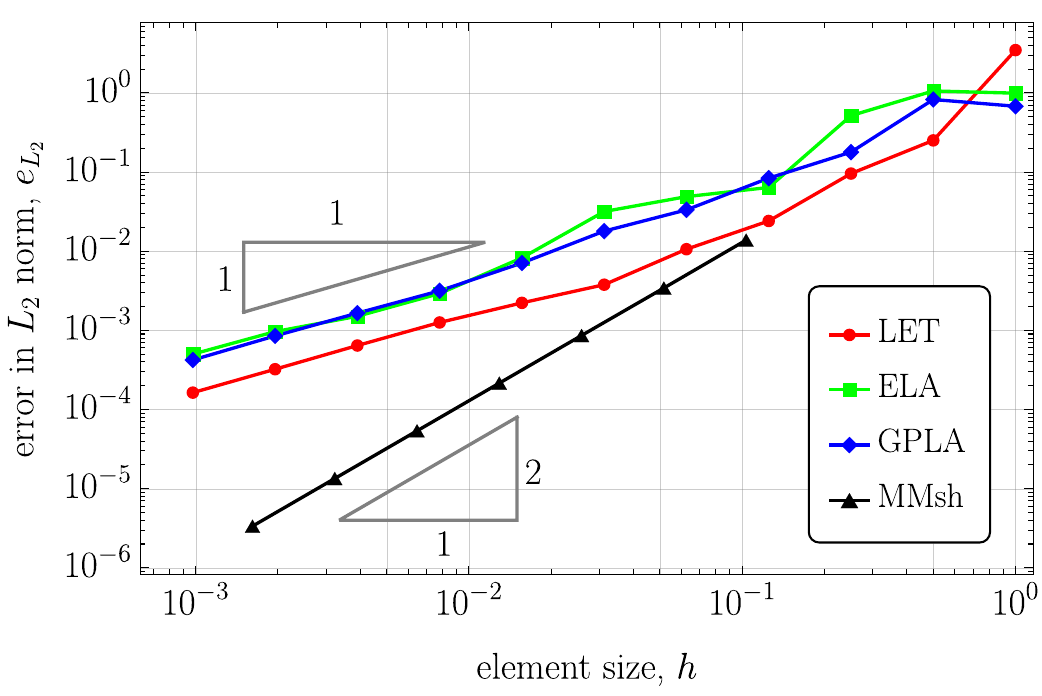} &
    \includegraphics[width=0.4\textwidth]{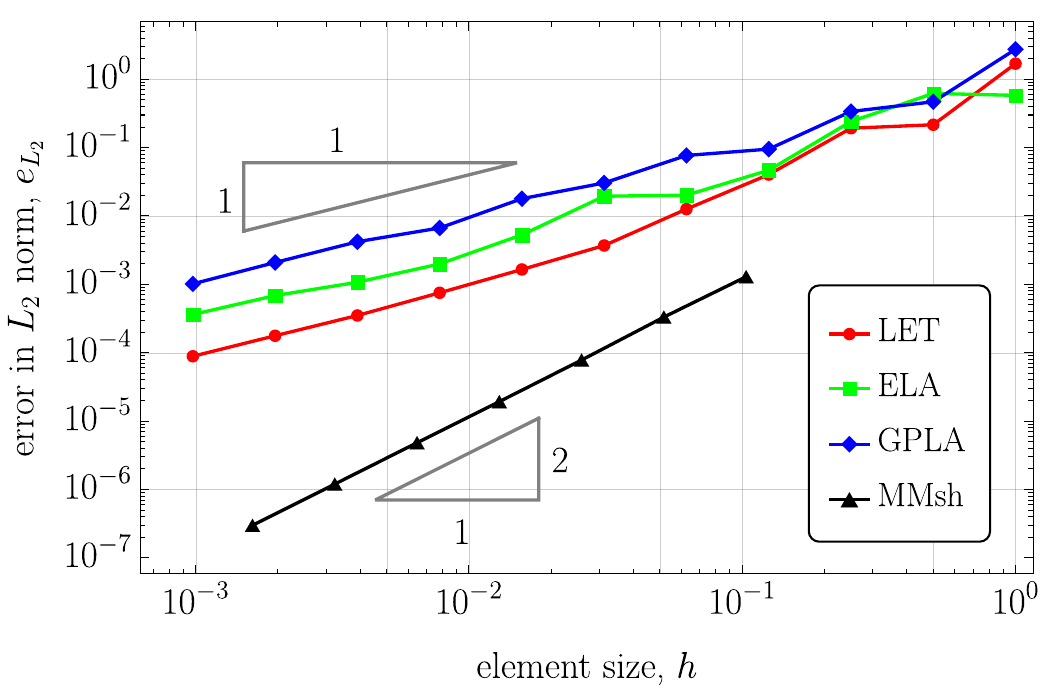} \\
    \hspace*{3em}{\footnotesize (a)} & \hspace*{3em}{\footnotesize (b)}
  \end{tabular}
  }
\caption{{Elastic inclusion problem: rate of convergence in $L_2$ norm for (a) soft inclusion ($E_2/E_1=10$) and (b) hard inclusion ($E_2/E_1=0.005$).}
}\label{inclL2}
\end{figure}

{Fig.~\ref{inclCR} shows the results obtained for LET, and for comparison, for ELA and GPLA, the two simple non-conforming mesh approaches that will be used as a reference in all subsequent examples. 
Fig.~\ref{inclCR} includes also the results obtained for a matching mesh as well as the results taken from the literature for exactly the same problem, specifically, for X-FEM \citep{Moes2003} (available only for the soft inclusion case) and for A-FEM \citep{Essongue2020}. 
It follows from Fig.~\ref{inclCR} that the convergence rate of LET, ELA, GPLA, and A-FEM is similar, approximately equal to $0.5$, but the error is the lowest for LET (in the hard inclusion case, the error of LET and A-FEM is similar). 
Since these methods employ a non-conforming approximation of the displacement field, the optimal convergence rate of 1, characteristic for a matching mesh and also for X-FEM, cannot be achieved. 
This is also illustrated in Fig.~\ref{inclL2} where the error in $L_2$ norm ($e_{L_2}=\|\bm{u}^h-\bm{u}^{\text{exact}}\|_{L_2(\Omega)}$) is shown. In this norm, the convergence rate is approximately equal to 1 for LET, ELA and GPLA, while it is equal to 2 for a matching mesh, as expected. 
It is stressed here that LET is not aimed to compete with more sophisticated methods, like X-FEM, in terms of accuracy. The advantage of LET is its simplicity and ease of implementation, and, at the same time, improved accuracy, as compared to ELA and GPLA.}

%
%
%
%

\begin{figure}[H]
\centerline{\includegraphics[width=0.4\textwidth]{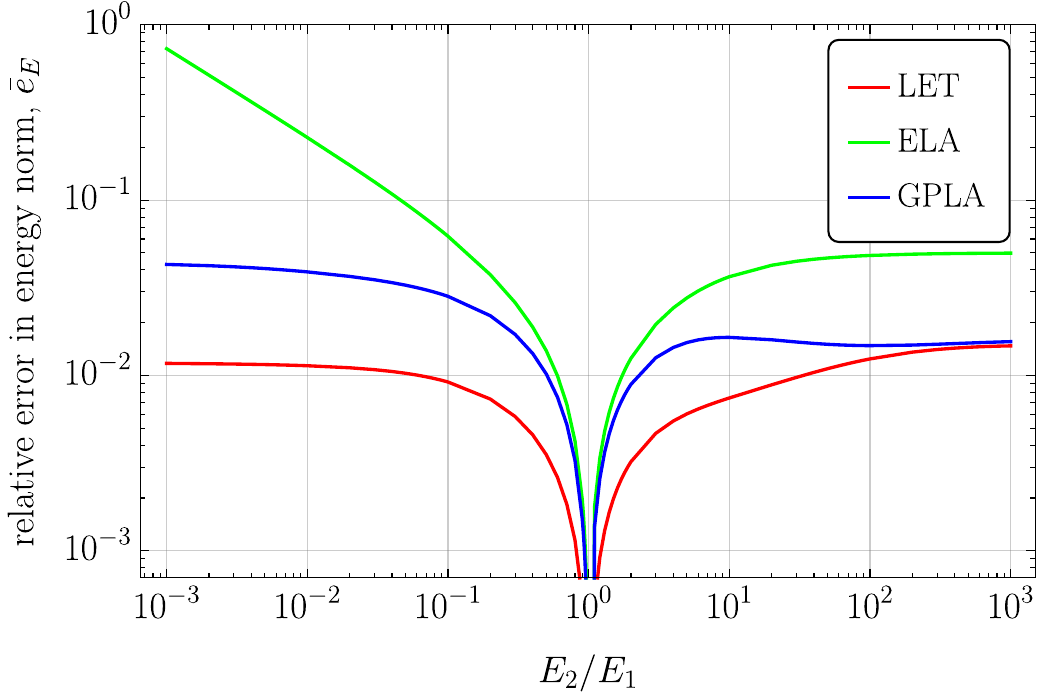}}
\caption{{Elastic inclusion problem: relative error in energy norm as a function of the Young's moduli contrast $E_2/E_1$.
}}\label{inclContrast}
\end{figure}

Fig.~\ref{inclContrast} shows the relative error {in energy norm} as a function of the Young's moduli contrast $E_2/E_1$ evaluated for $E_1=1$ and $\nu_1=\nu_2=0.25$, and for the element size $h=0.004$ ($500\times500$ elements). 
Again, the error is the lowest for LET, and the difference with respect to ELA and GPLA increases with increasing contrast, particularly in the case of hard inclusion ($E_2/E_1<1$).

%
%
%
%

%% file: sections/05_Eigen2DSSE.tex
\begin{figure}[H]
\centerline{
  \begin{tabular}{ccc}
    \hspace{0.5eM} \includegraphics[height=0.31\textwidth]{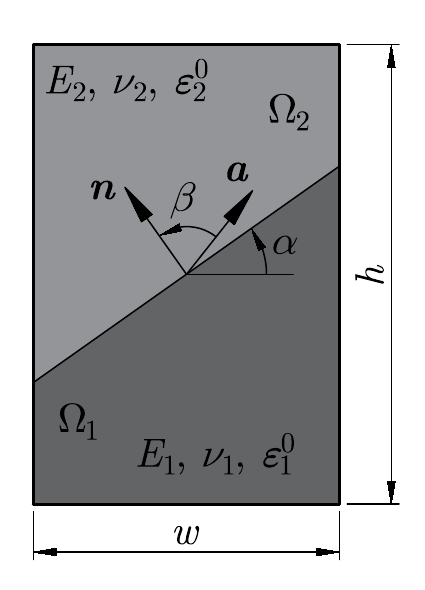} \hspace{0.5eM}  &
    \hspace{0.5eM}\raisebox{1.8em}[0pt][0pt]{\includegraphics[height=0.247\textwidth]{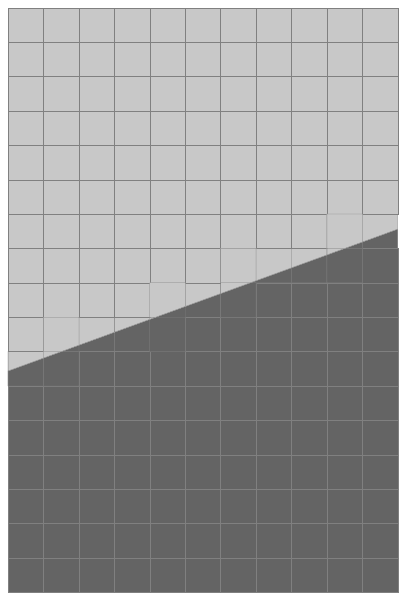}} \hspace{0.5eM} &
    \hspace{1.5eM}\raisebox{1.75em}[0pt][0pt]{\includegraphics[height=0.26\textwidth]{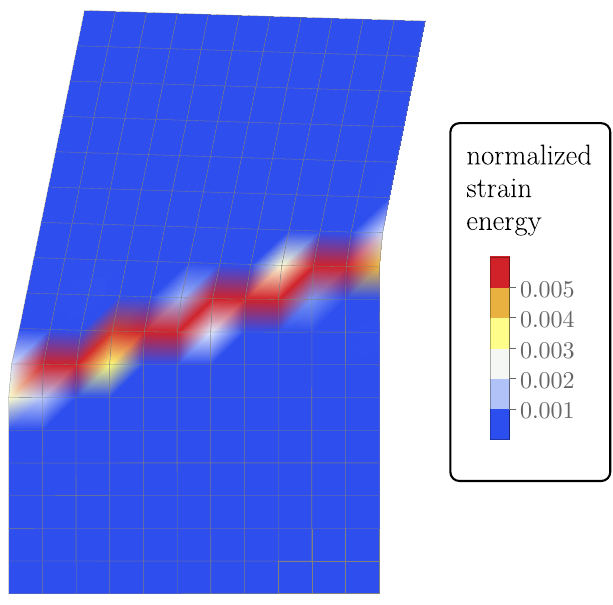}} \hspace{0.5eM} \\[-1ex]
    \hspace*{-1em}{\footnotesize (a)} & \hspace*{-0.3em}{\footnotesize (b)} & \hspace*{-1em}{\footnotesize (c)}
  \end{tabular}
  }
\caption{{Eigenstrain problem: (a) scheme of the problem; (b) computational domain with a coarse (non-conforming) mesh of $11\times17$ quadrilateral elements for a sample interface orientation $\alpha=\pi/9$ (the actual computations are carried out for a fine mesh of $101\times151$ elements); (c) deformed coarse mesh with the color map of the normalized elastic strain energy density, $W/(a^2 E^\ast)$, for $\beta=\pi/2$ (displacements are scaled for better visibility).
}}\label{eigenscheme}
\end{figure}

In this section we investigate the behaviour of LET for a 2D problem of elasticity with eigenstrain in the small-strain framework. 
Two elastic domains are separated by a planar interface inclined at an angle $\alpha$ to the horizontal axis, see Fig.~\ref{eigenscheme}(a). Each domain is homogeneous and is characterized by elastic constants $E_i$ and $\nu_i$ and by a homogeneous eigenstrain $\bm{\varepsilon}_i^0$ so that the elastic strain energy is a function of the elastic strain, $W_i=W_i(\bm{\varepsilon}^{\rm e})$, $\bm{\varepsilon}^{\rm e}=\bm{\varepsilon}-\bm{\varepsilon}_i^0$.

The eigenstrains in both phases are assumed compatible so that
\begin{equation}\label{eigen}
\Delta\bm{\varepsilon}=\bm{\varepsilon}_{2}^{0}-\bm{\varepsilon}_{1}^{0}=\frac{1}{2} \left( \bm{a}\otimes\bm{n} + \bm{n}\otimes\bm{a} \right) ,
\end{equation}
where $\bm{n}$ is a unit normal to the interface, $\bm{a}$ is a prescribed vector, {and to fix attention we assume that $\bm{\varepsilon}_{1}^{0}=\bm{0}$}. 
Accordingly, in the continuum setting, the total elastic strain energy vanishes,
\begin{equation}\label{zeroenergy}
\Int_{\Omega}W\left(\bm{\varepsilon}-\bm{\varepsilon}^{0}\right)\,\mathrm{d}\Omega=0 .
\end{equation}
For a non-conforming finite-element mesh, the local incompatibilities introduced by the discretization are accommodated by elastic strains, {see Fig.~\ref{eigenscheme}(c),} and the corresponding strain energy can be used as a measure of the error. 
The {normalized} error in energy norm is thus defined as
\begin{equation}\label{abserror}
{\tilde{e}_{E}
=\frac{1}{\left( a^2 E^\ast L \right)^{1/2}} \norm{\bm{u}^h-\bm{u}^{\text{exact}}}_{E(\Omega)}
=\left( \frac{1}{a^2 E^\ast L} \int_{\Omega} W(\bm{\varepsilon}^h-\bm{\varepsilon}^0)\,\mathrm{d}\Omega \right)^{1/2},}
\end{equation}
where the error is normalized by the interface length $L$ and also by {$a=\|\bm{a}\|$} and $E^\ast=\sqrt{E_1 E_2}$ so that the error does not depend on $a$ and depends on $E_1$ and $E_2$ only through their ratio, the contrast $E_2/E_1$.

%
%
%
%

The analysis is performed for a wide range of Young's moduli contrasts $E_2/E_1 \in(0.001,1000)$ with $\nu_1=\nu_2=0.25$ and for two values of the angle $\beta$ between vectors $\bm{a}$ and $\bm{n}$, namely $\beta=0$ and $\beta=\pi/2$. 
The actual computations are carried out for $E_1E_2=1$ and $a=1$. 
Dimensions of the rectangular domain are $w=10$ and $h=15$, and the domain is discretized into {a regular mesh of} $101\times151$ elements (element size $h\approx0.1$). 
An odd number of elements is adopted in each direction {and the interface passes through the centre of the domain so that, for all orientation angles,} the mesh is non-conforming (also for $\alpha=0$ and $\pi/2$). The boundaries are free, only the rigid body motion is prevented by enforcing adequate boundary conditions.

\begin{figure}[H]
\centerline{
  \begin{tabular}{cc}
    \includegraphics[width=0.4\textwidth]{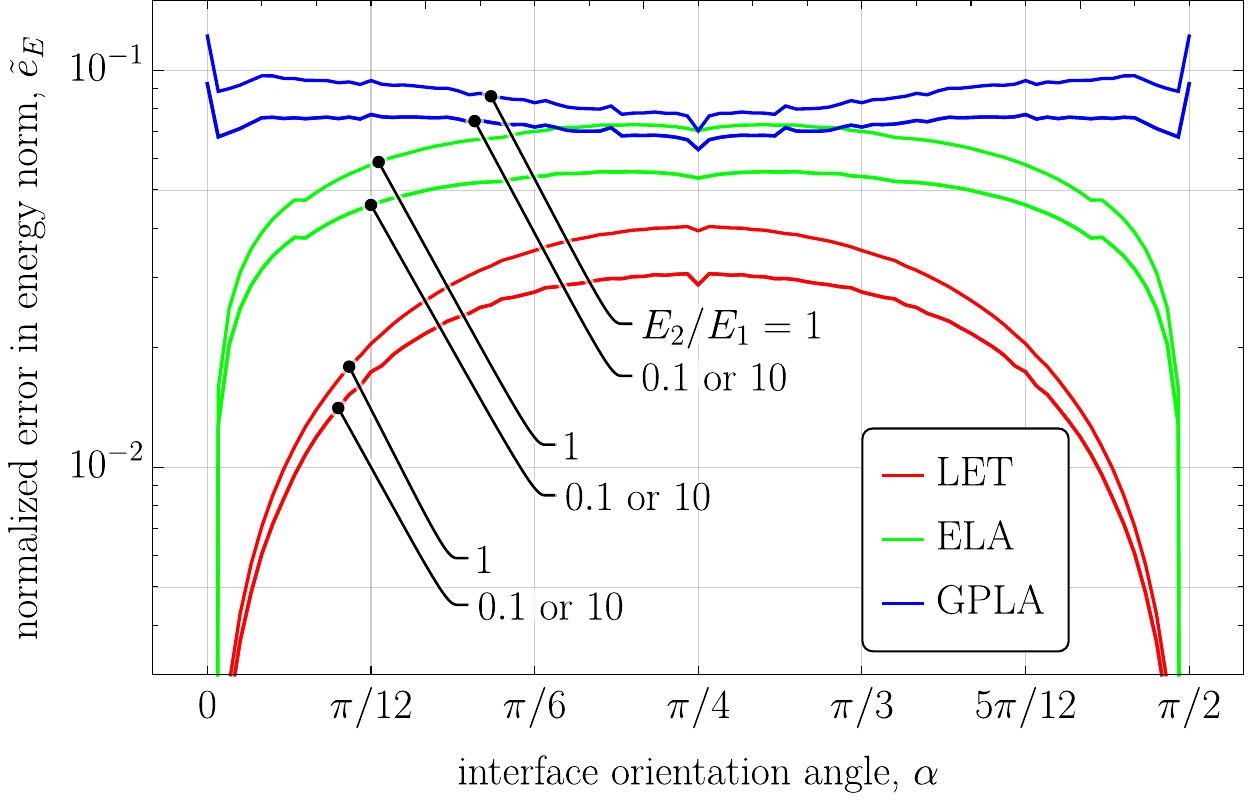} &
    \includegraphics[width=0.4\textwidth]{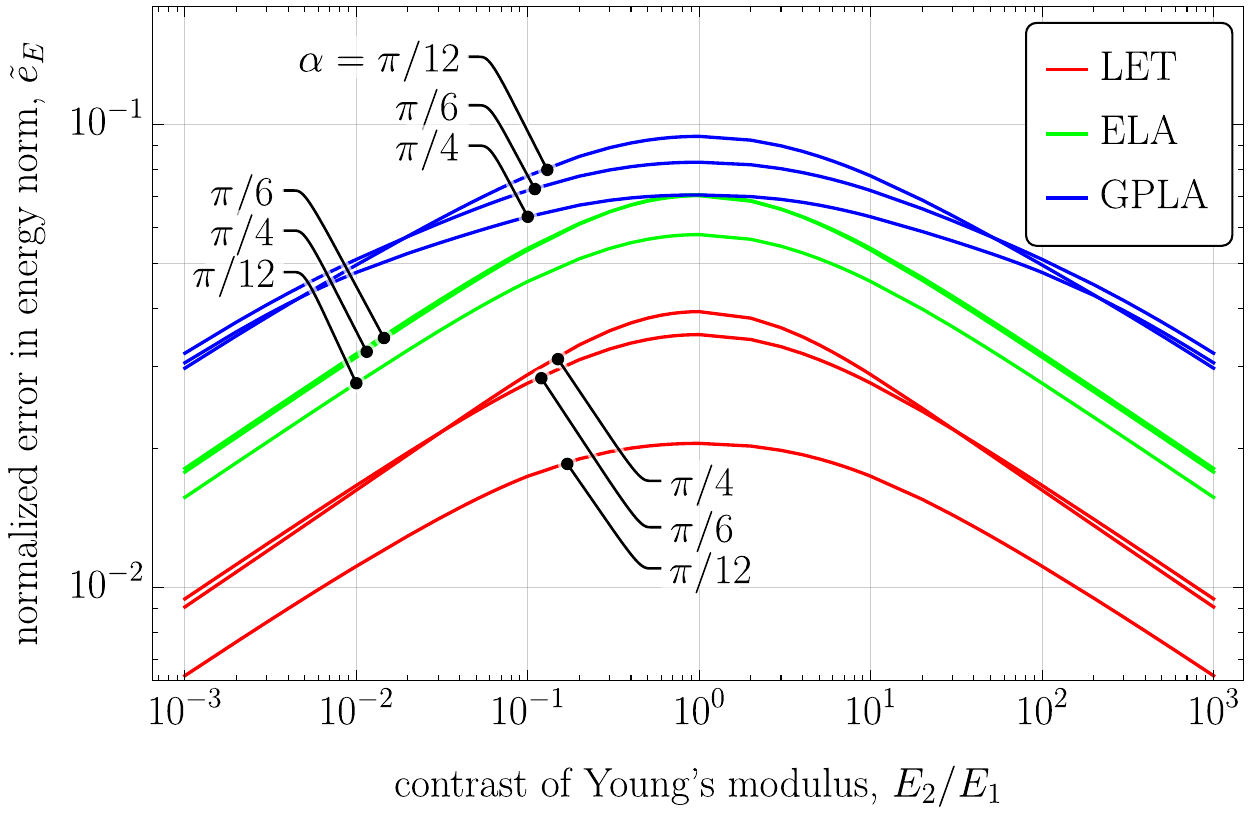} \\
    \hspace*{3em}{\footnotesize (a)} & \hspace*{3em}{\footnotesize (b)}
  \end{tabular}
  }
\caption{{Normalized} error in energy norm for $\beta=0$ as a function of: (a) the interface orientation {angle $\alpha$} and (b)~the Young's moduli contrast $E_2/E_1$.
}\label{eigen:c0n}
\end{figure}

\begin{figure}[H]
\centerline{
  \begin{tabular}{cc}
    \raisebox{0.11eM}[0pt][0pt]{
    \includegraphics[width=0.4\textwidth]{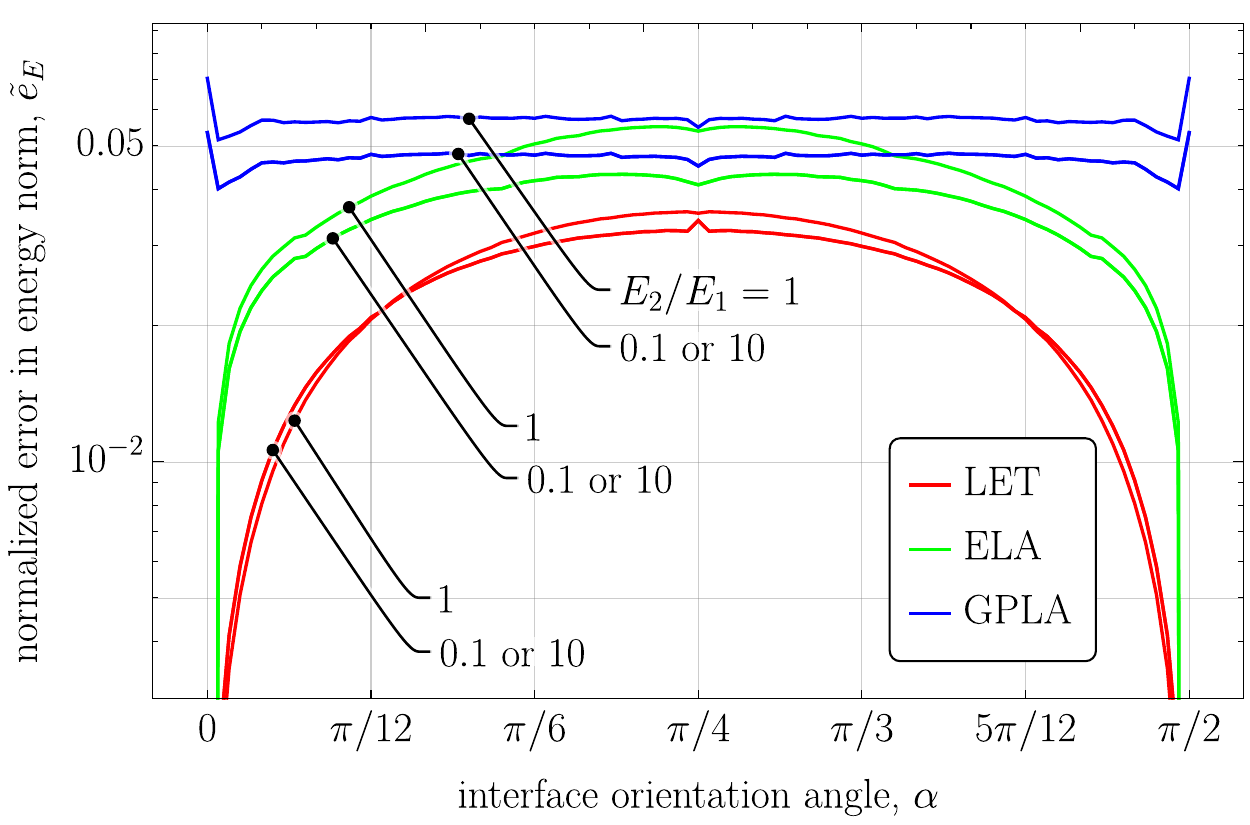}
    } &
    \includegraphics[width=0.4\textwidth]{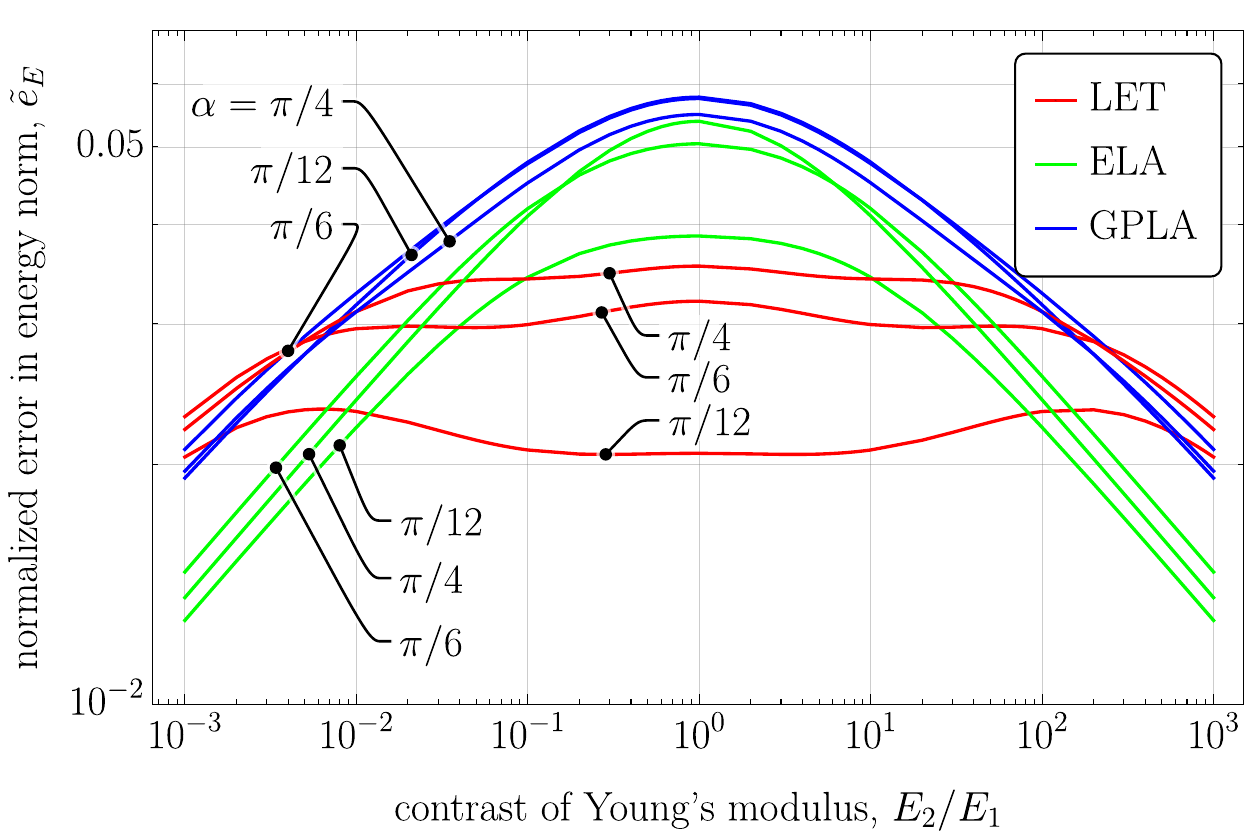} \\
    \hspace*{3em}{\footnotesize (a)} & \hspace*{3em}{\footnotesize (b)}
  \end{tabular}
  }
\caption{{Normalized} error in energy norm for $\beta=\pi/2$ as a function of: (a) the interface orientation {angle $\alpha$} and (b) the Young's moduli contrast $E_2/E_1$.}\label{eigen:c90n}
\end{figure}

Representative results are shown in Fig.~\ref{eigen:c0n} for $\beta=0$ and in Fig.~\ref{eigen:c90n} for $\beta=\pi/2$. 
{Intermediate values of $\beta$ are not considered since the corresponding solutions can be obtained by the superposition of those for $\beta=0$ and $\beta=\pi/2$ (even if the error, as a nonlinear function of the solution, cannot be obtained by superposition).}


Figs.~\ref{eigen:c0n}(a) and \ref{eigen:c90n}(a) show the dependence of the error on the interface orientation angle $\alpha$. 
As expected, the individual diagrams exhibit symmetry with respect to $\alpha=\pi/4$. 
Likewise, the diagrams in Figs.~\ref{eigen:c0n}(b) and \ref{eigen:c90n}(b), which depict the dependence on the contrast $E_2/E_1$ exhibit symmetry with respect to $E_2/E_1=1$ (recall that $\nu_1=\nu_2$ and $E_1 E_2=1$). 
It is also seen that, for LET and ELA, the error vanishes for $\alpha=0$ and $\pi/2$, i.e., when the interface is parallel to element edges. 

%
%
%
%
%
%
%
%

Figs.~\ref{eigen:c0n} and~\ref{eigen:c90n} show that in most cases the error is the lowest for LET. 
However, for $\beta=\pi/2$, when the eigenstrain jump is a shear strain, LET performs better than ELA and GPLA only for moderate contrasts, see Fig.~\ref{eigen:c90n}(b).

%% file: sections/06_Inclusion3DSSE.tex
One of the advantages of LET over ELA and GPLA is its ability to adapt to continuous changes in the position of the interface within a single finite element. 
{In LET, if} the position of the interface is varied in a continuous {manner, the} volume fractions of the phases and the {orientation} of the interface also change in a continuous manner, whereas in ELA and GPLA these changes are taken into account in a step-wise manner (volume fraction) or not at all ({orientation} of the interface). 
This effect is illustrated here by considering a 3D cubic cell of dimension $L$ with a central inclusion of varying diameter $D$.

\begin{figure}[h]
    \centering
    \begin{subfigure}[b]{0.49\textwidth}
        \centering
        \includegraphics[width=0.7\textwidth, trim={0 6pt 0 0},clip]{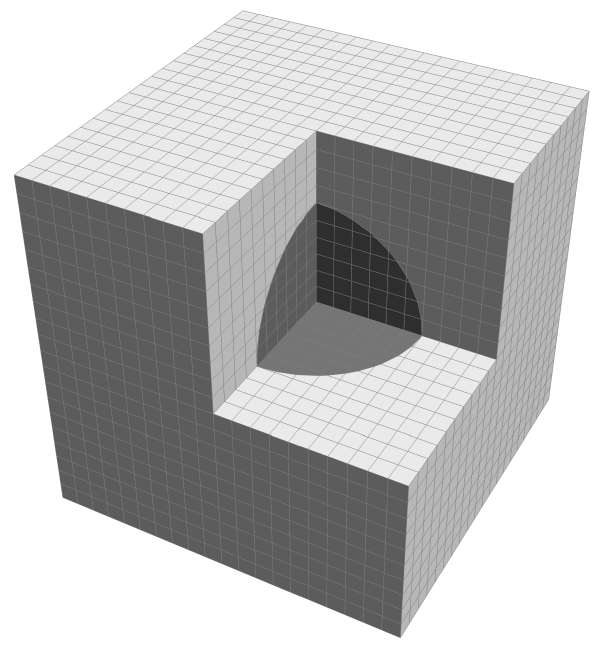}
        \caption{}
        \label{3dinclusionFEM06}
    \end{subfigure}
    \begin{subfigure}[b]{0.49\textwidth}
        \centering
        \includegraphics[width=0.7\textwidth]{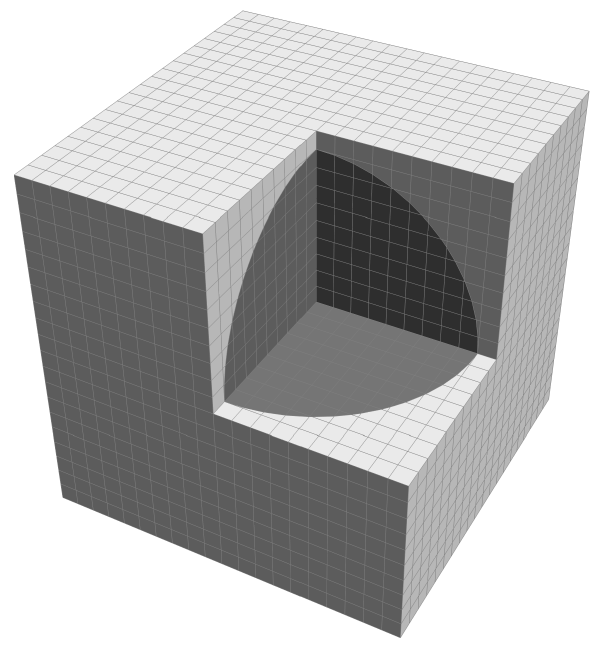}
        \caption{}
        \label{3dinclusionFEM09}
    \end{subfigure}
    \caption{Periodic unit cell with a spherical elastic inclusion of the diameter that varies between $D/L=0.6$ (a) and $D/L=0.9$ (b). A fixed regular finite-element mesh ($20\times20\times20$ elements) is used.}
    \label{3dinclusionFEM}
\end{figure}

Both phases are linear elastic with the properties specified as $E_1=10$, $\nu_1=0.3$ (inclusion) and $E_2=1$, $\nu_2=0.2$ (matrix). 
The inclusion diameter is varied between $D/L=0.6$ and $D/L=0.9$, and a regular 
{mesh of $20\times20\times20$ elements} is used, see Fig.~\ref{3dinclusionFEM}. 
Periodic boundary conditions are enforced and the overall elastic moduli tensor is determined in a standard manner {by subjecting the unit cell to 6 linearly independent macroscopic strains (actually 3 are sufficient due to symmetry). The overall elastic moduli tensor is then determined in terms of the resulting overall stress tensors.} 
Below, the results are reported in terms of the directional Young's modulus $E_{100}=(S_{1111})^{-1}$, where $S_{ijkl}$ denotes the components of the elastic compliance tensor.

%
%
%
%

\begin{figure}[h]
\centerline{
  \begin{tabular}{cc}
    \includegraphics[width=0.4\textwidth]{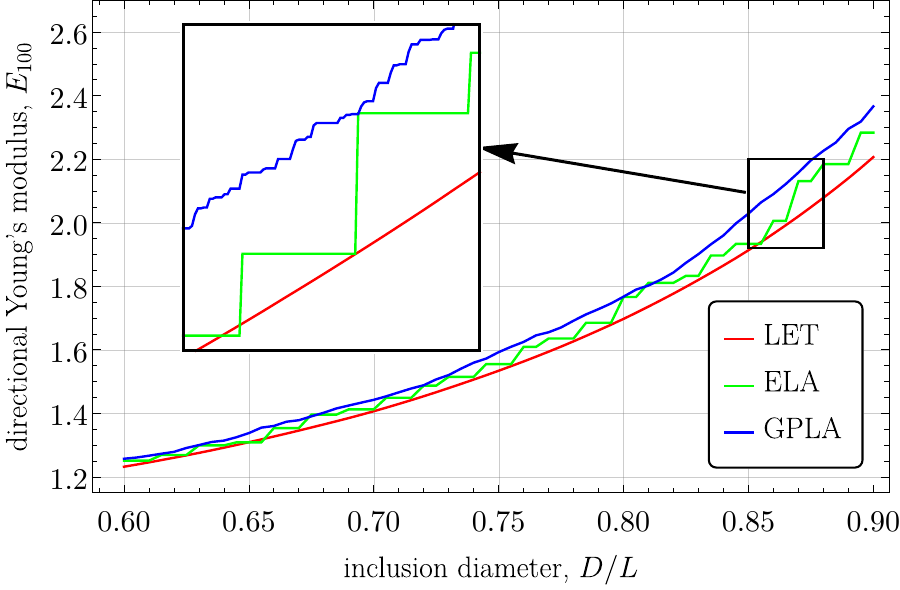} &
    \includegraphics[width=0.4\textwidth]{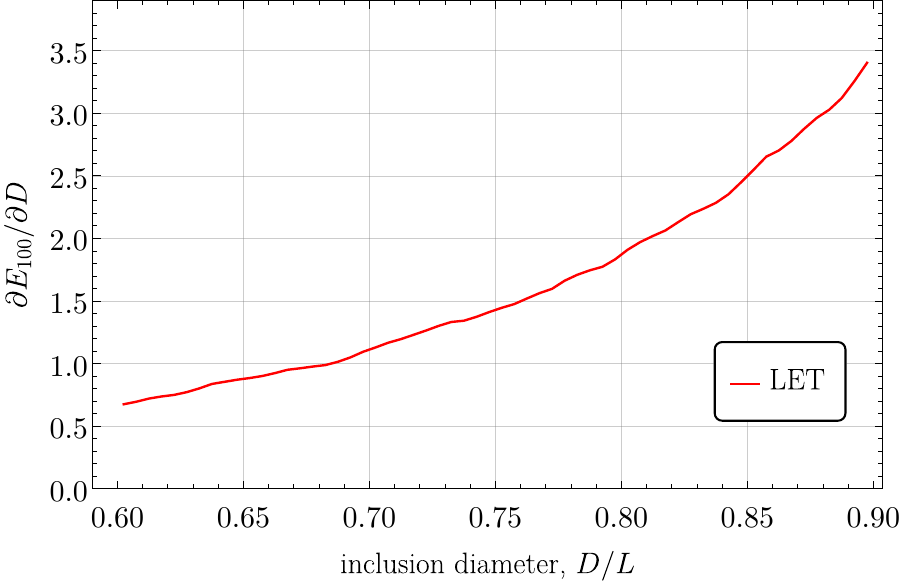} \\
    \hspace*{2em}{\footnotesize (a)} & \hspace*{2em}{\footnotesize (b)}
  \end{tabular}
  }
\caption{(a) Dependence of the directional Young's modulus, $E_{100}$, on the inclusion diameter $D$. The inset shows the results computed with a finer step so that the jumps are clearly visible for both ELA and GPLA. (b) Dependence of the derivative of the directional Young's modulus, {$\partial E_{100}/\partial D$}, on the inclusion diameter $D$.
}\label{3dinclusion2}
\end{figure}

{Fig.~\ref{3dinclusion2}(a)} shows the dependence of $E_{100}$ on the inclusion diameter. 
The mesh is here rather coarse, hence the predictions of the three methods (LET, ELA, GPLA) differ visibly. 
However, the important difference is that the dependence is smooth in the case of LET, while in the case of ELA and GPLA the overall properties change in a step-wise manner, see the inset in {Fig.~\ref{3dinclusion2}(a)}. Here, the overall moduli (e.g., $E_{100}$) exhibit a jump whenever the element ({for} ELA) or Gauss point ({for} GPLA) is assigned to a different phase when the interface position is changed. 
Clearly, LET is free of such artifacts, and the overall moduli depend on the inclusion diameter in a continuous fashion. 

%
%
%
%

For completeness, the derivative of the dependence of $E_{100}$ on {$D$}, as predicted by LET, is shown {Fig.~\ref{3dinclusion2}(b)}. The derivative is here computed using the finite difference scheme in terms of two subsequent data points. The small irregularities that can be seen in {Fig.~\ref{3dinclusion2}(b)} result from the error introduced by LET. 
It follows that the response is continuous and its derivative is meaningful, which suggests that LET can be considered as a candidate for treating moving interface problems, such as microstructure evolution or shape optimization.

%% file: sections/07_Wovencell.tex
\begin{figure}[h]
    \centering
    \begin{subfigure}[b]{0.49\textwidth} 
        \centering
        \includegraphics[width=0.8\textwidth]{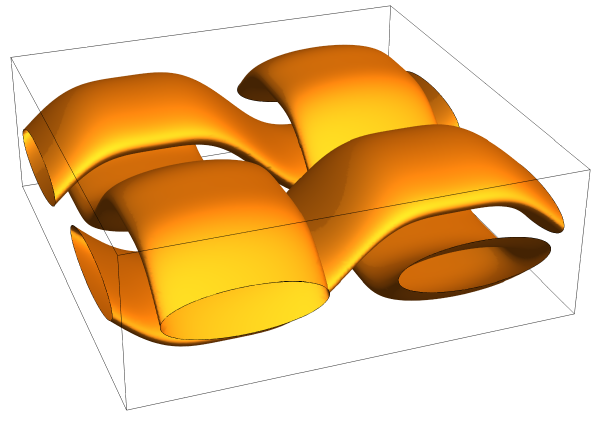}
        \caption{}
        \label{woven.geometry}
    \end{subfigure}
    \hfill
    \begin{subfigure}[b]{0.49\textwidth}
        \centering
        \includegraphics[width=0.8\textwidth]{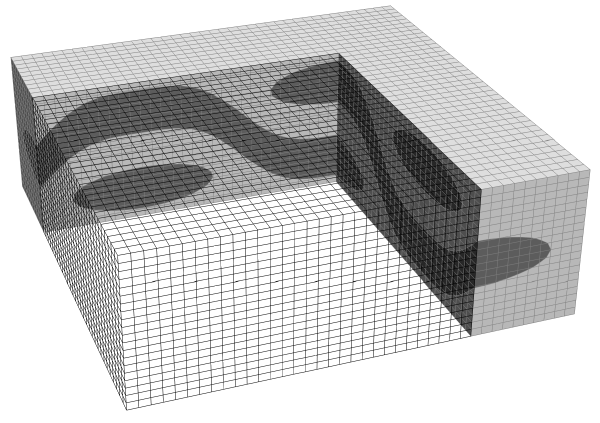}
        \caption{}
        \label{woven.model}
    \end{subfigure}
    \caption{Periodic woven-cell problem: (a) geometry, (b) finite-element mesh ($40\times40\times20$ elements).
    \label{woven}
    }
\end{figure}

In this example, unlike the previous ones, the internal geometry is more complex, and for this reason the real advantage of LET over conforming-mesh discretization can be appreciated. 
The model consists of a 3D periodic cell of {the} dimensions $L \times L \times H = 2 \times 2 \times 0.7$, in which four interlaced fibres are immersed in the matrix, Fig.~\ref{woven}.

%
%
%
%

For a coarse mesh, it may happen that one element is cut by two interfaces. Accordingly, as commented in Section~\ref{LEM}, a separate level-set function is introduced for each fibre. If an element is cut by two interfaces, so that it contains portions of the matrix and two fibres, the {total volume fraction of the fibres is simply taken as} the sum of the volume fractions of the individual fibres and the lamination orientation is determined by averaging those determined individually for each {interface}.

The geometry of the fibres aligned with the $x_2$-axis is defined by the centreline $(\pm x_1^0,x_2,\mp x_3^0(x_2))$ parameterized by $x_2$ and by the elliptical cross-section (in the $(x_1,x_3)$-plane) specified by the following inequality
%
%
\begin{equation}
    \sqrt{ \left( \frac{x_1 \pm x_1^0}{a} \right)^2 + \left( \frac{x_3 \pm x_3^0(x_2)}{b} \right) ^2 }-1 \leqslant 0,
\end{equation}
%
%
where $a=0.35$ and $b=0.11$ are the semi-major and semi-minor axes of the ellipse, $x_1^0=L/4$ defines the offset between the fibres in the $x_1$-direction, function $x_3^0(y)$ is specified as
%
%
\begin{equation}
    x_3^0(x_2) = A \left( \frac{9}{8} \sin{ \left( \frac{2\pi x_2}{L} \right)}+\frac{1}{8}\sin{ \left( \frac{6\pi x_2}{L} \right)} \right) ,
\end{equation}
%
%
and $A=0.2$ is the amplitude of the function $x_3^0(x_2)$. The origin of the coordinate system is located at the centre of the unit cell. 
The geometry of the fibres aligned with the $x_1$-axis is defined analogously. 

The finite-deformation framework is employed, and both the matrix and the fibres are assumed to be hyperelastic, characterized by a compressible neo-Hookean strain energy function. The elastic properties are specified as $E_1=100$ (fibres), $E_2=1$ (matrix), and $\nu_1=\nu_2=0.45$. 

Periodic boundary conditions are imposed and loading is applied by prescribing the overall deformation gradient $\bar{\bm{F}}$. Three deformation modes are considered, namely isochoric tension and two cases of simple shear. 
The isochoric tension along the $x_1$-axis is specified by
%
%
\begin{equation}
    \bar{\bm{F}} = ( 1+\epsilon ) \bm{e}_1 \otimes \bm{e}_1 + \frac{1}{\sqrt{1+\epsilon}} \left( \bm{e}_2 \otimes \bm{e}_2 + \bm{e}_3 \otimes \bm{e}_3 \right) ,
\end{equation}
%
%
where $\epsilon$ denotes the elongation, and $\bm{e}_i$ are the orthonormal basis vectors. 
The simple shear is specified by
%
%
\begin{equation}
    \label{eq:shear}
    \bar{\bm{F}} = \bm{I} + \gamma \, \bm{s} \otimes \bm{n} ,
\end{equation}
%
%
with $\bm{s}=\bm{e}_1$ and $\bm{n}=\bm{e}_2$ (case \#1) and $\bm{s}=\frac{1}{\sqrt{2}}(\bm{e}_1+\bm{e}_2)$ and $\bm{n}=\frac{1}{\sqrt{2}}(\bm{e}_2-\bm{e}_1)$ (case \#2). 

\begin{figure}
    \centerline{\scriptsize
        \begin{tabular}{ccc}
            \includegraphics[height=3.5cm]{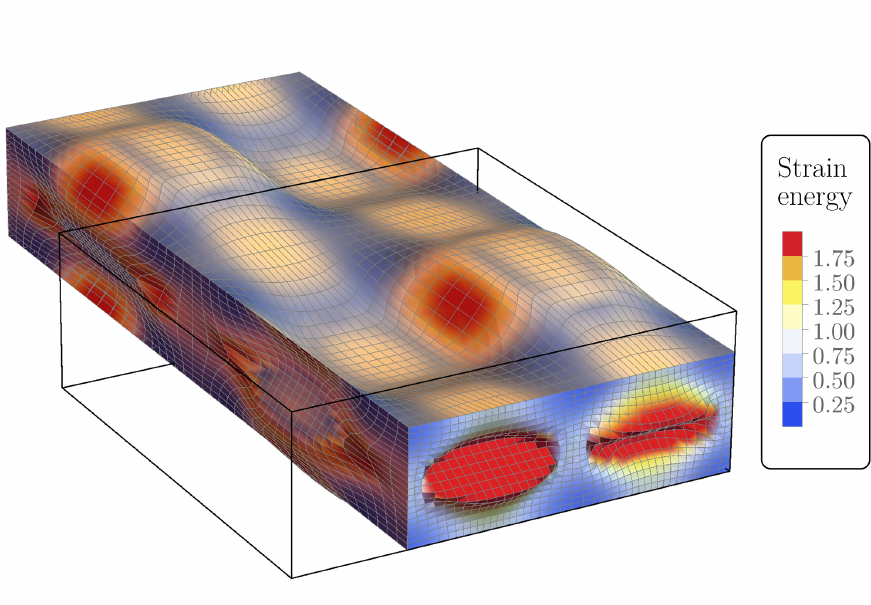} &
            \includegraphics[height=3.5cm]{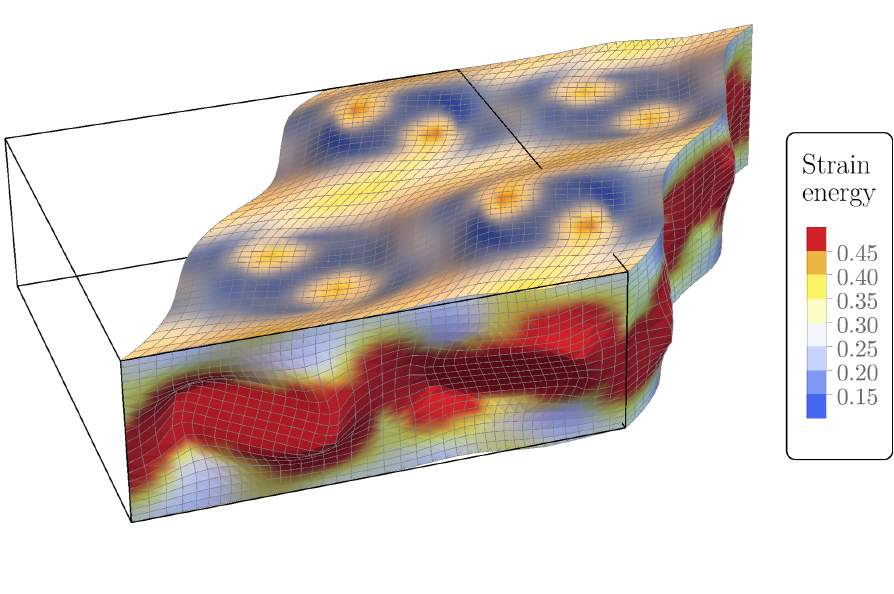} &
            \includegraphics[height=3.5cm]{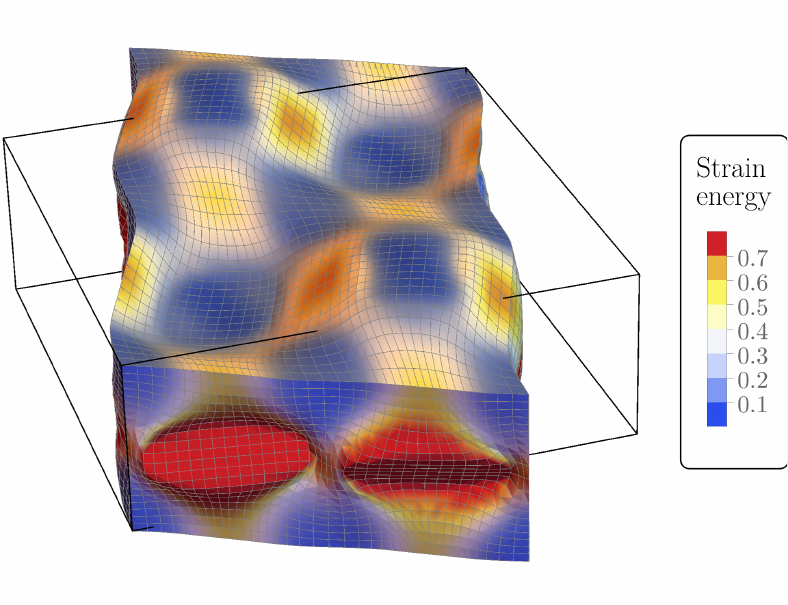} \\
            (a) & (b) & (c)
        \end{tabular}
        }
    \caption{Woven cell: deformed configuration for (a) isochoric tension ($\epsilon=0.8$), (b) simple shear (case \#1, $\gamma=0.7$), and (c) simple shear (case \#2, $\gamma=0.7$). Colour maps show the elastic strain energy density $W$.}
    \label{woven.defomation}
\end{figure}

In the convergence studies reported below, a family of regular meshes of hexahedral elements is used with 10 to 40 elements along the $x_1$- and $x_2$-directions (element size $h$ varied between 0.2 and 0.05) and with 5 to 20 elements in the $x_3$-direction, respectively. 
{As a reference, the results obtained for a fine mesh of $80\times80\times40$ elements ($h=0.025$) are used, and both LET and ELA are employed for this purpose (the two methods give very similar results; for ELA, in the simple shear case \#2, the solution could not be achieved at the maximum load due to convergence problems).} 
The F-bar formulation is employed to avoid volumetric locking effects \citep{deSouzaNeto1996}.
Fig.~\ref{woven.defomation} illustrates the three deformation modes for the mesh of {$40\times40\times20$ elements}.

%
%
%
%

\begin{figure}
    \centerline{\scriptsize
        \begin{tabular}{ccc}
            \includegraphics[width=0.35\textwidth]{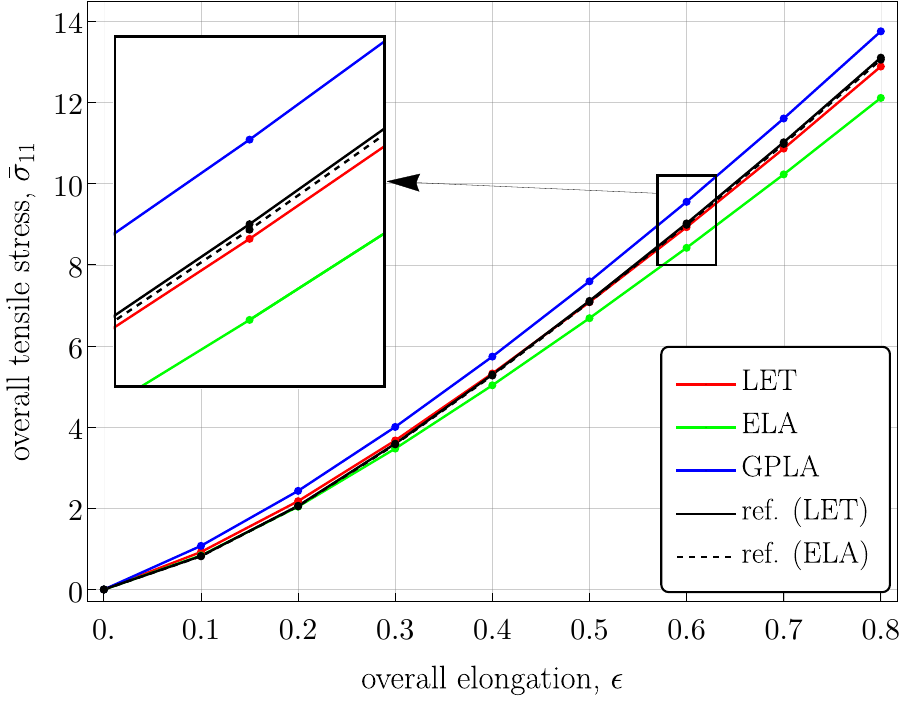} &
            \includegraphics[width=0.35\textwidth]{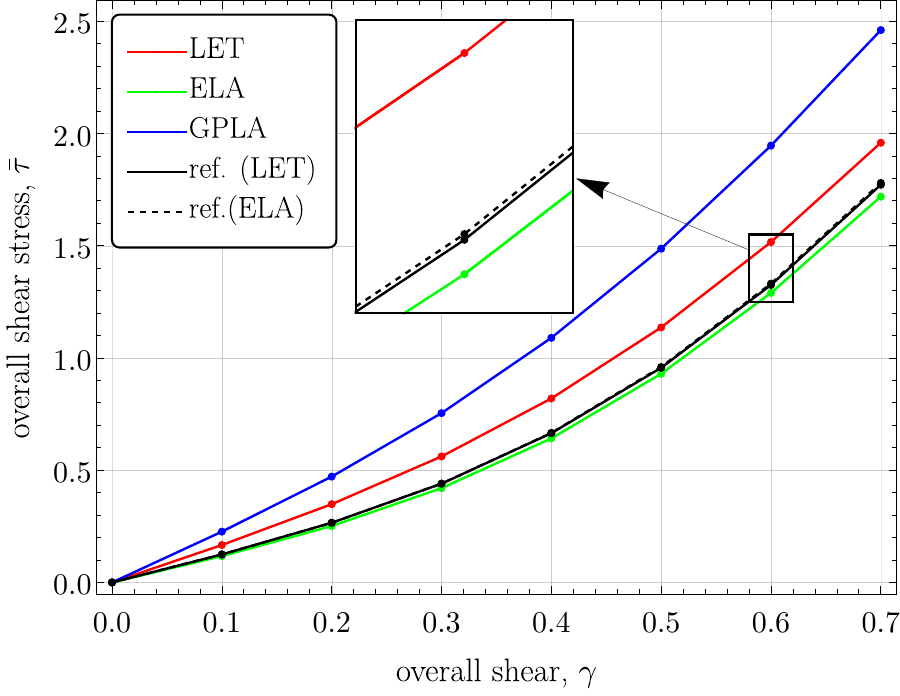} &
            \includegraphics[width=0.35\textwidth]{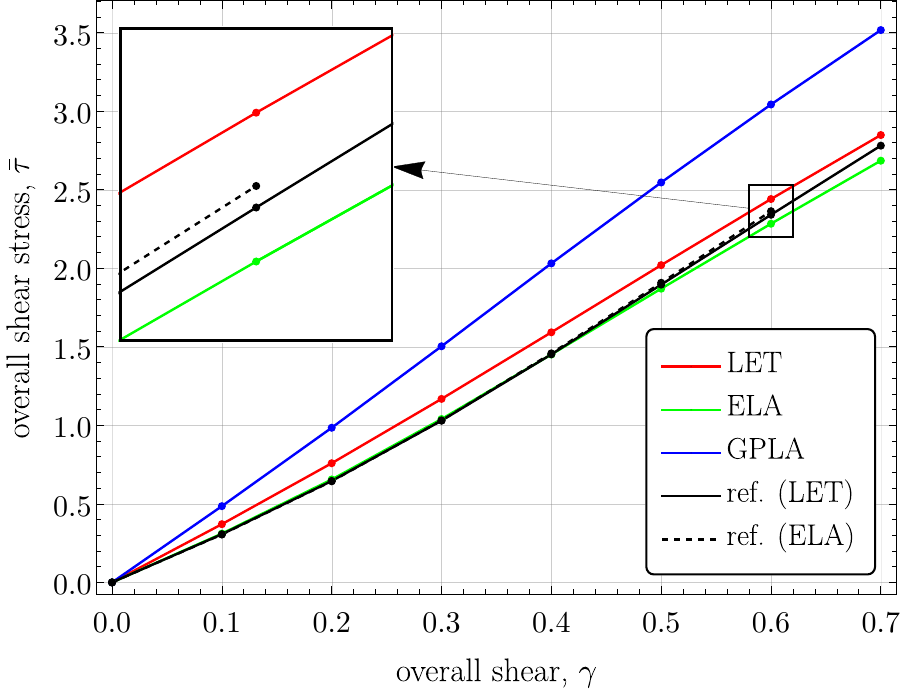} \\
            (a) & (b) & (c)
        \end{tabular}
        }
    \caption{Woven cell: overall stress--strain response for (a) isochoric tension, (b) simple shear (case \#1), and (c) simple shear (case \#2). The results correspond to the mesh of $20\times20\times10$ elements (element size $h=0.1$). As a reference, the LET {and ELA} results obtained for a fine mesh ($80\times80\times40$ elements, $h=0.025$) are used.}
    \label{woven.mech-resp-C100}
\end{figure}

The overall stress--strain response predicted using the $20\times20\times10$ mesh is shown in Fig.~\ref{woven.mech-resp-C100}. 
In the case of isochoric tension, the $\bar{\sigma}_{11}$ component of the overall Cauchy stress $\bar{\bm{\sigma}}$ is shown as a function of the overall elongation $\epsilon$, Fig.~\ref{woven.mech-resp-C100}(a). 
In the case of simple shear, the shear stress $\bar{\tau}=\bm{s}\cdot\bar{\bm{\sigma}}\cdot\bm{n}$ is shown as a function of the overall shear $\gamma$, Fig.~\ref{woven.mech-resp-C100}(b,c). 
Results obtained for a four times finer mesh are included in Fig.~\ref{woven.mech-resp-C100} as a reference. 

%
%
%
%

In Fig.~\ref{woven.mech-resp-C100}, the mesh is relatively coarse, hence the visible differences between the three methods (LET, ELA, GPLA). 
In all cases, GPLA delivers the stiffest response, with the largest error with respect to the reference results. 
The remaining two methods (LET and ELA) deliver similar results that agree well with the reference ones, except for simple-shear case \#1, where the LET results are visibly stiffer. 
Accordingly, for this specific mesh density ($h=0.1$), ELA seems to perform the best. 
However, this conclusion does not apply to other mesh densities, as illustrated below.

Fig.~\ref{woven.ConvRates} shows the {overall} stress at the maximum strain as a function of the mesh size $h$. 
It follows that predictions of LET are stable (i.e., reasonably close to the reference solution) over the entire range of mesh densities studied. 
Likewise, GPLA delivers stable results, although with a significantly higher error. 
On the other hand, ELA performs badly for coarser meshes, which is associated with a poor representation of the internal geometry by ELA.

%
%
\begin{figure}
    \centerline{\scriptsize
        \begin{tabular}{ccc}
            \includegraphics[width=0.35\textwidth]{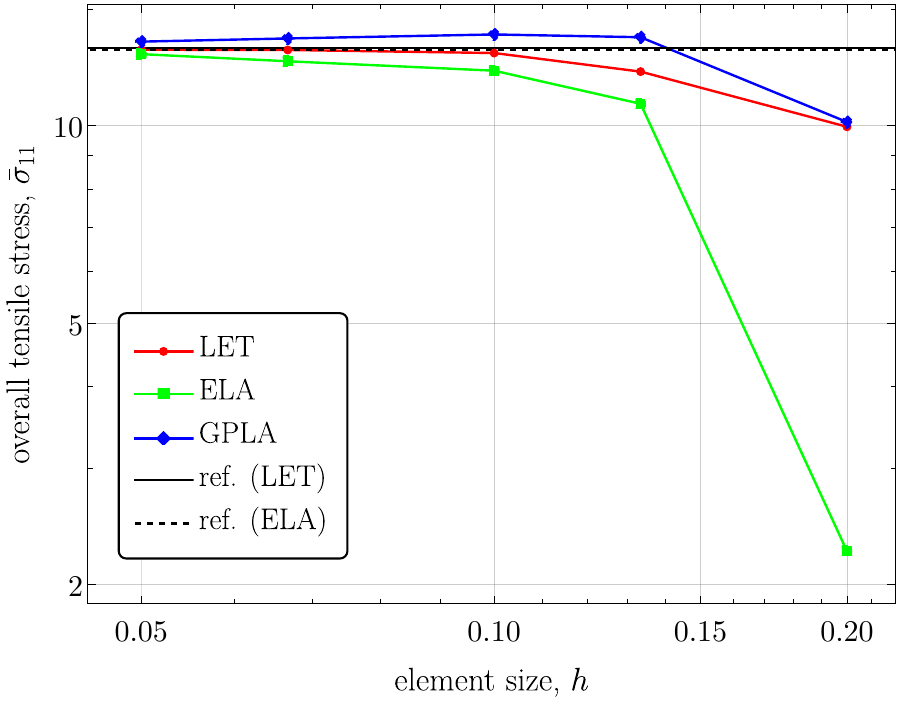} &
            \includegraphics[width=0.35\textwidth]{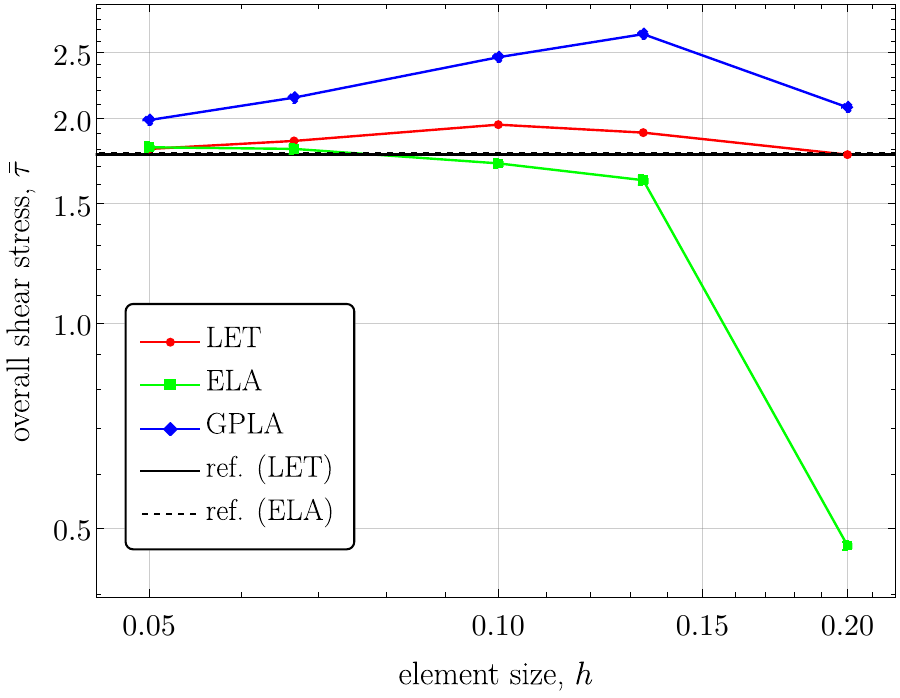} &
            \includegraphics[width=0.35\textwidth]{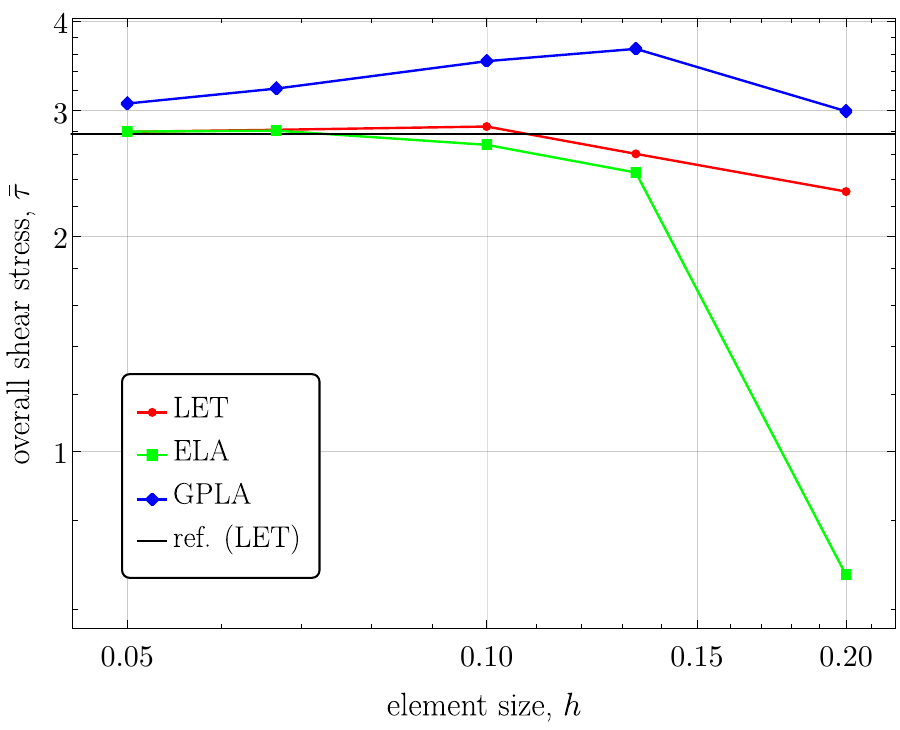} \\
            (a) & (b) & (c)
        \end{tabular}
        }
    \caption{Woven cell: convergence of the {overall} stress (at the maximum strain, see Fig.~\ref{woven.mech-resp-C100}) with element size $h$ for (a) isochoric tension, (b) simple shear (case \#1), and (c) simple shear (case \#2). As a reference, the LET {and ELA} results obtained for a fine mesh ($80\times80\times40$ elements, $h=0.025$) are used.}
    \label{woven.ConvRates}
\end{figure}
%
%

%% file: sections/08_Inclusion2DFSEP.tex
In this example, we consider a 2D periodic unit cell with a circular inclusion. Both phases are elastic-plastic and plane-strain conditions are assumed. 
In the continuum setting, the position of the inclusion within the unit cell is arbitrary in view of periodicity, and it does not affect the overall response. 
This is not {the} case in the discrete setting when the position of the inclusion with respect to the finite-element mesh is an additional geometric feature that may affect the response, as revealed by our preliminary studies. 
Accordingly, in this section we examine this effect in detail. 
Specifically, the overall response under simple shear is studied for 100 randomly selected positions of the inclusion and for a family of regular meshes of $N\times N$ elements with $N=2,4,8,\ldots,256$. 
Below, for each method considered (LET, ELA, GPLA), the responses obtained for a given mesh density are averaged and compared to the reference (``exact'') solution obtained using a high-resolution conforming mesh involving over 3 million elements. The standard deviation is also examined as an indicator of the sensitivity of the response to the position of the inclusion.

The geometric and material parameters adopted in this example are the following. 
The dimensions of the unit cell are $L\times L$ with $L=2$, and the inclusion radius is $R=0.6$. 
The finite-deformation framework is adopted and both phases are governed by the finite-strain $J_2$ plasticity model with linear isotropic hardening, see \ref{app:ep} for more details. The yield stress $\sigma_{\rm y}$ is thus specified by {$\sigma_{\rm y}^{}(\alpha)=\sigma_{\rm y}^0+K\alpha$}, where $\alpha$ denotes the accumulated plastic strain. 
The elastic properties of the matrix and inclusion are the same, $E=70000$, $\nu=0.25$, and so is the hardening modulus $K=2000$. The initial yield stress of the inclusion, $\sigma_{{\rm y},1}^0=70$, is lower than that of the matrix, $\sigma_{{\rm y},2}^0=120$, which induces an inhomogeneous deformation within the unit cell {once plastic deformation occurs}.

\begin{figure}[H]
    \centerline{\scriptsize
        \begin{tabular}{cccc}
            \includegraphics[height=0.22\textwidth]{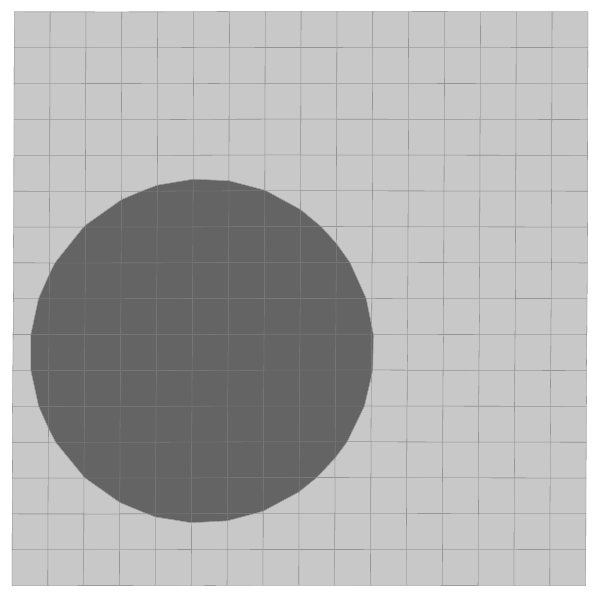} &
            \includegraphics[height=0.22\textwidth]{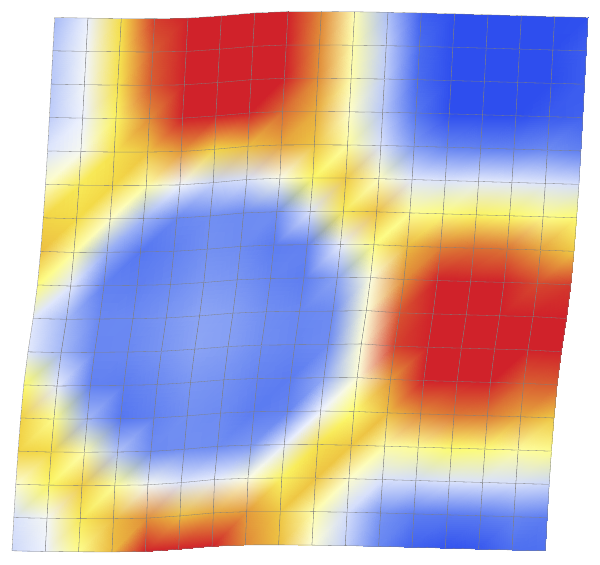} &
            \includegraphics[height=0.22\textwidth]{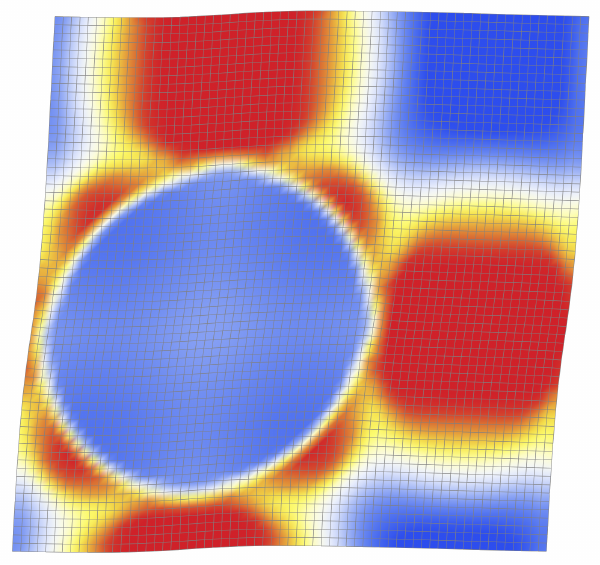} &
            \includegraphics[height=0.22\textwidth]{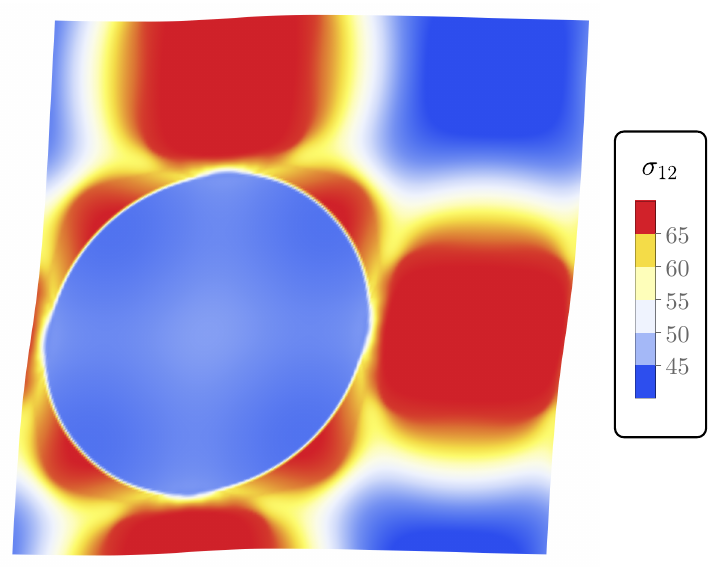} \\[1ex]
            (a) & (b) & (c) & (d)\hspace*{3em}
        \end{tabular}
        }
    \caption{Elasto-plastic composite: (a) $16 \times 16$ mesh with a randomly positioned circular inclusion; (b,c,d) deformed mesh (displacements scaled 20 times) with the distribution of the shear component $\sigma_{12}$ of the Cauchy stress tensor for $16\times16$ (b), $64\times64$ (c) and $256\times256$ (d) mesh.
    }
    \label{fig:EPinclusion}
\end{figure}

The unit cell is loaded in simple shear by prescribing the overall deformation gradient $\bar{\bm{F}}$ according to Eq.~\eqref{eq:shear} with $\bm{s}=\bm{e}_1$ and $\bm{n}=\bm{e}_2$. The initial stage of deformation is considered with the overall shear $\gamma$ increasing from 0 to 0.004 so that the details of the elastic-to-plastic transition are revealed. 
{A sample finite-element mesh with the inclusion at a sample position within the unit cell} is shown in Fig.~\ref{fig:EPinclusion}(a), and the deformation mode along with the shear component $\sigma_{12}$ of the Cauchy stress tensor are shown in Fig.~\ref{fig:EPinclusion}(b-d) for three selected mesh densities.

%
%
%
%

\begin{figure}[H]
    \centerline{\scriptsize
        \begin{tabular}{ccc}
            \includegraphics[height=0.26\textwidth]{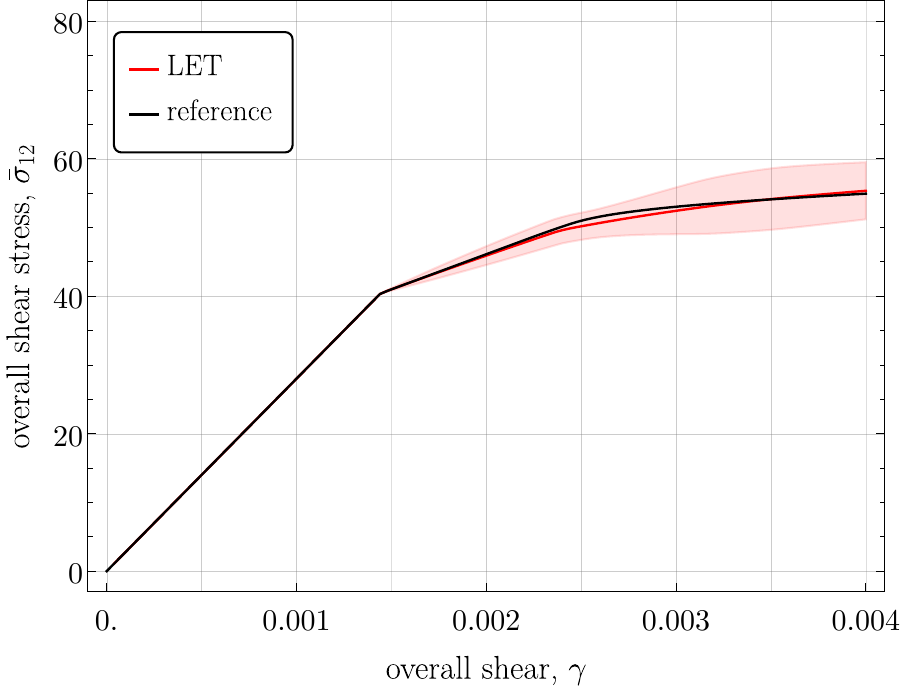} &
            \includegraphics[height=0.26\textwidth]{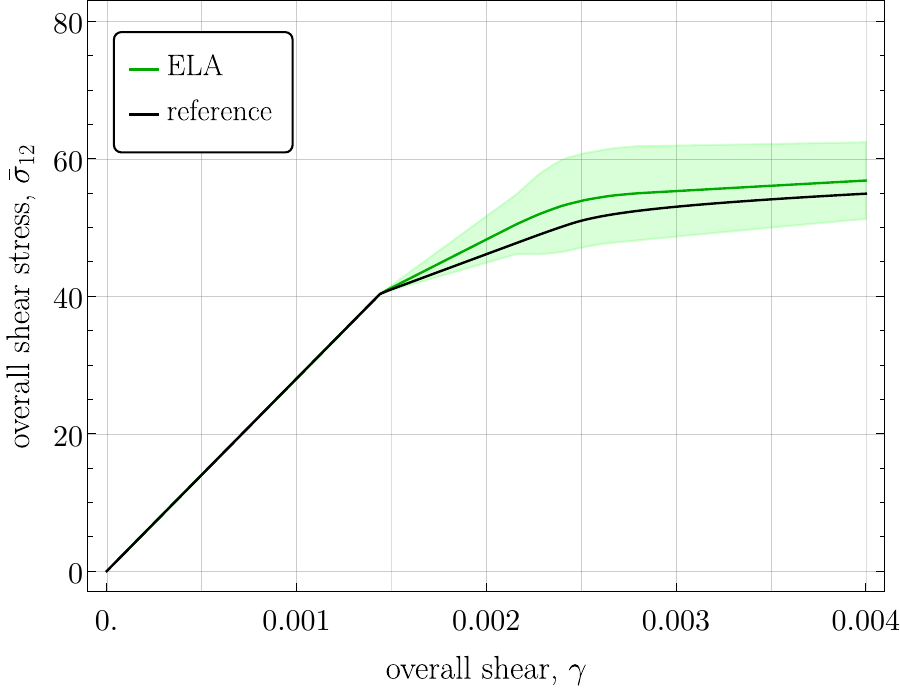} &
            \includegraphics[height=0.26\textwidth]{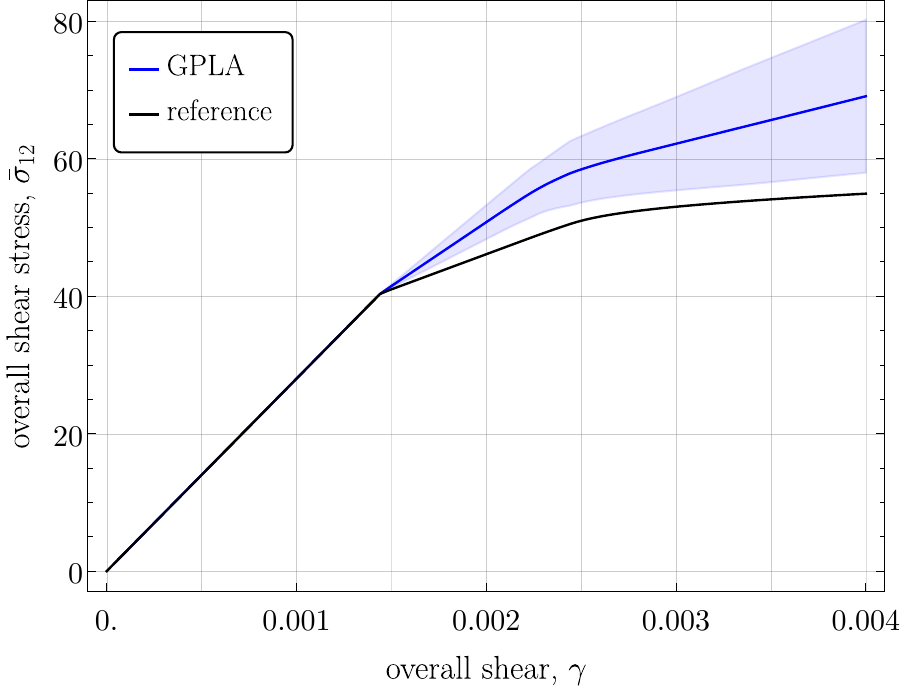} \\[1ex]
            (a) & (b) & (c)
        \end{tabular}
        }
    \caption{Elasto-plastic composite: overall stress--strain ($\bar{\sigma}_{12}$--$\gamma$) response obtained for the $4\times4$ mesh (element size $h=0.5$) and for LET (a), ELA (b) and GPLA (c). In each case, the average over 100 random inclusion positions is indicated by a solid line and the corresponding shaded area indicates the spread ($\pm3$ standard deviations).}
    \label{fig:EPinc:4x4}
\end{figure}

Fig.~\ref{fig:EPinc:4x4} shows the overall stress--strain response ($\bar{\sigma}_{12}$ component of the overall Cauchy stress as a function of the overall shear $\gamma$) for a coarse mesh of $4\times4$ elements. 
Here, the average response is compared to the reference solution and, moreover, the shaded area represents the spread of the individual responses corresponding to the randomly positioned inclusions (the width of the shaded area is set equal to $\pm3$ standard deviations). 
Since the mesh is here coarse ($4\times4$ elements), visible differences with respect to the reference solution are apparent. 
It can be seen that LET delivers the most accurate results in terms of both the average and the spread. 
The accuracy is visibly worse in the case of ELA and significantly worse in the case of GPLA.

%
%
%
%

\begin{figure}[H]
    \centerline{\scriptsize
        \begin{tabular}{cc}
            \includegraphics[width=0.4\textwidth]{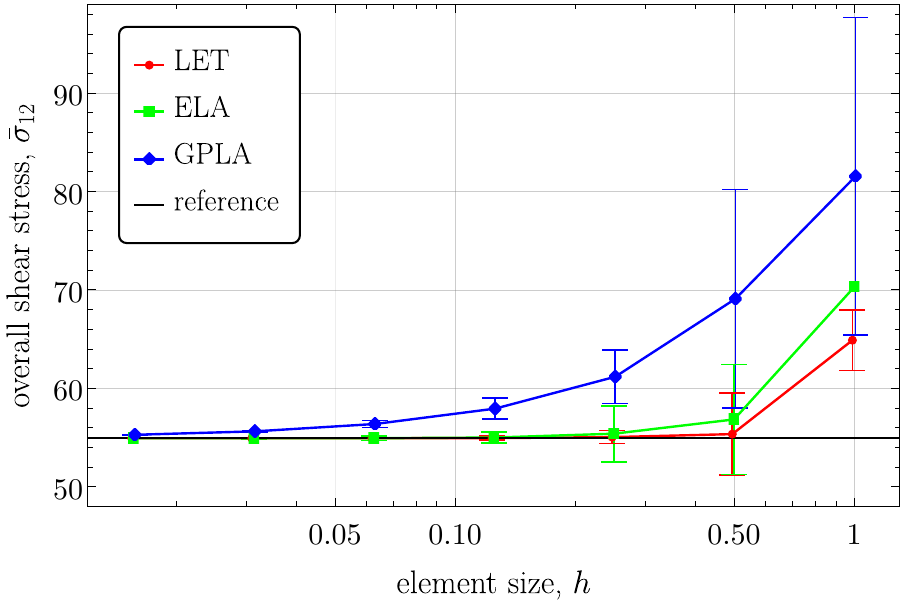} &
            \includegraphics[width=0.4\textwidth]{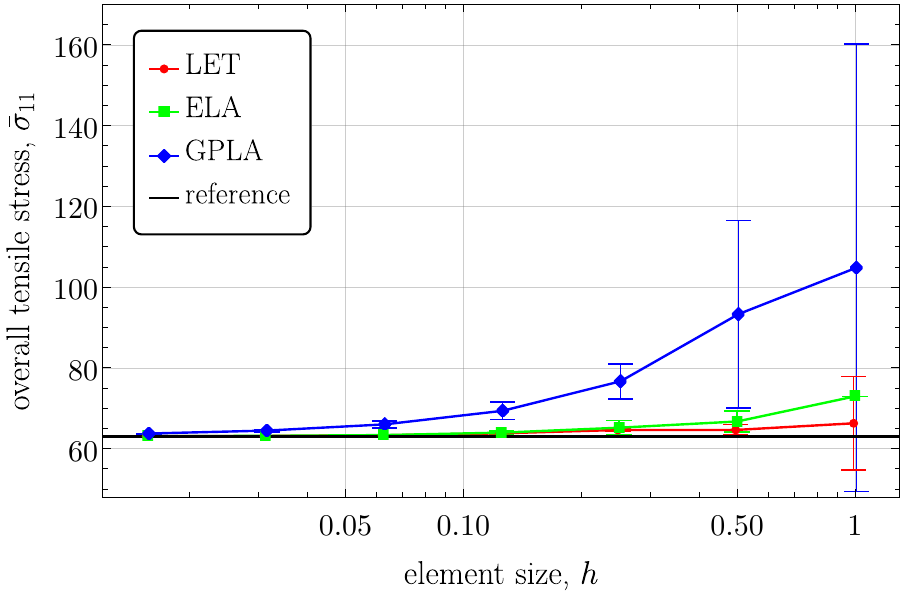} \\[1ex]
            \hspace*{3em}(a) & \hspace*{3em}(b)
        \end{tabular}
        }
    \caption{Elasto-plastic composite: dependence of the {average overall stress on the element size $h$ for (a) simple shear (shown is the shear stress at $\gamma=0.004$) and (b) isochoric tension (shown is the tensile stress at the elongation of 0.004).} The error bars indicate the spread ($\pm3$ standard deviations).}
    \label{fig:EPinc:convergence}
\end{figure}

Convergence of the results with mesh refinement is illustrated in Fig.~\ref{fig:EPinc:convergence}. 
{This figure, in addition to simple shear, includes also the results corresponding to isochoric tension (i.e., pure shear).}
In the case of LET and ELA, the averaged stress converges quickly to the reference value, LET converging somewhat faster. 
However, in the case of LET, the spread vanishes significantly faster than in the case of ELA. 
Consistent with the other results, the accuracy of GPLA is the worse. 

%
%
%
%

Note that in the case of ELA, the results obtained for the coarsest mesh considered ($2\times2$ elements, $h=1$) exhibit no spread, see Fig.~\ref{fig:EPinc:convergence}. This is because in this case all elements are assigned to the matrix phase regardless of the position of the inclusion (and the unit cell is thus homogeneous).

%% file: sections/99_AppendixA.tex
\section{Incremental computational scheme for an elastic-plastic simple laminate}
\label{app:ep}

In this appendix, we discuss the case of a simple laminate composed of two elastic-plastic materials. 
As discussed below, the corresponding computational model involves a nested iterative-subiterative Newton scheme. Its consistent linearization is crucial so that the Newton method can be effectively used on the structural level. In this work, this is achieved by using the automatic differentiation (AD) technique that is available in \emph{AceGen} \citep{Korelc2009,KorelcWriggers2016}, and below we use the compact AD-based notation introduced in \citep{Korelc2009}.

The incremental constitutive equations of finite-strain plasticity are rather standard \citep{SimoHughes1998}, and the details are omitted here. The specific AD-based formulation of elastoplasticity which is adopted here follows that developed in \citep{Korelc2009}, see also \citep{KorelcStupkiewicz2014,KorelcWriggers2016}. 
On the other hand, the treatment of the laminated microstructure is based on that developed in \citep{Sadowski2017} for the incremental Mori--Tanaka scheme (with due differences). The corresponding AD-based formulation is provided below in the form of a pseudocode with only short comments, while for the details the reader is referred to \citep{Sadowski2017}.

The AD-based notation employed in the pseudocodes below uses a special notation to denote the computational derivative, i.e., the derivative evaluated by AD. 
The computational derivative is denoted by $\hat{\delta}f/\hat{\delta}\brm{a}$, where $f$ is a function defined by an algorithm (or computer program) in terms of independent variables collected in vector $\brm{a}$. 
The actual dependencies present in the algorithm can be overridden or modified by introducing the so-called AD exceptions that are denoted by a vertical bar following the derivative with additional specifications in the subscript. 
The details can be found in \citep{Korelc2009,KorelcWriggers2016}.

Adopting the finite-strain framework, the elastic strain energy of phase $i$ is expressed as a function of the deformation gradient $\bm{F}_i=\bm{F}_i^{n+1}$ and the vector $\brm{h}_i=\brm{h}_i^{n+1}$ of internal (history) variables at the current time step $t=t_{n+1}$,
\begin{equation}
W_i=W_i(\bm{F}_i,\brm{h}_i) ,
\end{equation}
and the Piola stress $\bm{P}_i=\bm{P}_i^{n+1}$ is thus given by
\begin{equation}
\label{eq:Pi:ep}
\bm{P}_i=\frac{\partial W_i(\bm{F}_i,\brm{h}_i)}{\partial\bm{F}_i} .
\end{equation}
Here and below, the superscript $n+1$ denoting the quantities at $t_{n+1}$ is omitted to make the notation more compact. Time-discrete evolution of the internal variables $\brm{h}_i$ is governed by a set of nonlinear equations written symbolically in the residual form as
\begin{equation}
\label{eq:local}
\brm{Q}_i(\bm{F}_i,\brm{h}_i,\brm{h}_i^n)=\brm{0} ,
\end{equation}
where $\brm{h}_i^n$ denotes the known internal variables at the previous time step $t=t_n$. In computational plasticity, the incremental equations of elastoplasticity are usually solved using the return-mapping algorithm, which leads to the \emph{state update algorithm} that is outlined in Algorithm~\ref{alg:1} using the AD-based notation. In this algorithm, the local problem~\eqref{eq:local} is solved iteratively using the Newton method, and the derivative of the implicit dependence of the solution $\brm{h}_i$ on $\bm{F}_i$ (denoted by $\brm{G}_i$ in Algorithm~\ref{alg:1}) is computed in the standard manner \citep{Michaleris1994,Korelc2009},
\begin{equation}
\label{eq:implicit}
\frac{\partial\brm{h}_i}{\partial\bm{F}_i}=-\left(\frac{\partial\brm{Q}_i}{\partial\brm{h}_i}\right)^{-1}\frac{\partial\brm{Q}_i}{\partial\bm{F}_i} .
\end{equation}

\begin{algorithm}
\caption{\newline\texttt{StateUpdate[ ]}: state update algorithm for phase $i$}
\label{alg:1}
\begin{spacing}{1.2}
\begin{algorithmic}
 \State {\bf input:} $\bm{F}_i$, $\brm{h}_i^n$
 \State $\phi_i^{\text{trial}} \gets \phi_i(\bm{F}_i, \brm{h}_i^n)$
 \If {$\phi_i^{\text{trial}}<0$}
 \State $\brm{h}_i  \gets \brm{h}_i^n$ \vspace{0.5ex}
 \State $\brm{G}_i \gets \brm{0}$ \vspace{0.5ex}
 \Else
 \State $\brm{h}_i  \gets \brm{h}_i^n$
 \Repeat
 \State $\brm{A}_i \gets \dfrac{\hat{\delta} \brm{Q}_i \left( \bm{F}_i, \brm{h}_i, \brm{h}_i^n \right)}{\hat{\delta} \brm{h}_i}$
      \vspace{0.5ex}
      \Comment{tangent matrix, $\brm{A}_i=\displaystyle\frac{\partial\brm{Q}_i}{\partial\brm{h}_i}$}
 \State $\Delta \brm{h}_i \gets -\brm{A}_i^{-1} \brm{Q}_i$ \vspace{0.5ex}
 \State $\brm{h}_i \gets \brm{h}_i + \Delta \brm{h}_i$
 \Until $\| \Delta \brm{h}_i\| \le tol$\vspace{0.5ex}
 \State $\brm{G}_i \gets -\brm{A}_i^{-1} \left. \dfrac{\hat{\delta} \brm{Q}_i}{\hat{\delta} \bm{F}_i}
   \right|_{\brm{h}_i = {\rm const}}$
   \Comment{$\brm{G}_i=\displaystyle\frac{\partial\brm{h}_i}{\partial\bm{F}_i}$}
 \EndIf
 \State $\brm{h}_i \gets \brm{h}_i \Big|_{\gfrac{\RD \brm{h}_i}{\RD \bm{F}_i} 
    = \brm{G}_i}$ 
    \Comment{introduce the implicit dependence of $\brm{h}_i$ on $\bm{F}_i$}
    \vspace{0.5ex}
 \State $\bm{P}_i \gets \left. \dfrac{\hat{\delta} W_i ( \bm{F}_i,
    \brm{h}_i )}{\hat{\delta} \bm{F}_i} \right|_{\brm{h}_i={\rm const}}$
    \Comment{AD exception ensures that $\bm{P}_i$ is computed correctly}
    \vspace{0.5ex}
 \State {\bf return:} $\brm{h}_i$, $\bm{P}_i$, $\brm{G}_i$
\end{algorithmic}
\end{spacing}
\end{algorithm}

The specific constitutive functions of the finite-strain {$J_2$} plasticity model used in this work, see Section~\ref{sec:EPcomp}, are summarized in Box~\ref{box:Constitutive}, see the formulation I-C-C-b in Box~1 in \citep{KorelcStupkiewicz2014}. Box~\ref{box:Constitutive} provides also the corresponding definitions of the vector of internal variables $\brm{h}_i$ and of the local residual $\brm{Q}_i$. For brevity, the subscript $i$ is omitted in Box~\ref{box:Constitutive}.

\begin{mybox}
\caption{Constitutive equations of finite-strain $J_2$ plasticity with isotropic hardening. {Phase index $i$ is omitted for brevity.}
\label{box:Constitutive}}
\vspace{0ex}
	\centerline{
    \begin{tabular}{l}
    \hline
		Given: $\bm{F}, \bm{C}_{{\rm p},n}^{-1}, \gamma_n$ \qquad
		Find: $\bm{C}_{{\rm p}}^{-1}, \gamma$ \rule[-1.2ex]{0em}{4.2ex} \\ \hline
		$\bm{b}_{\rm e}=\bm{F} \bm{C}_{{\rm p}}^{-1} \bm{F}^{\rm T}$ \rule{0em}{3ex} \\
		$I_1 = \tr{\bm{b}_{\rm e}}, \quad I_3 = \det{\bm{b}_{\rm e}}$ \rule{0em}{3ex} \\
		$W = \frac{1}{2}\mu \left( I_1-3-\log{I_3} \right) + \frac{1}{4}\lambda \left( I_3-1-\log{I_3} \right)$ \rule{0em}{3ex} \\[0.ex]
      	$\bm{\tau}=2 \bm{b}_{\rm e} \dfrac{\partial W}{\partial \bm{b}_{\rm e}}$ \hspace*{5.65cm} $\triangleright\;\text{automation: } \bm{\tau} \leftarrow 2\bm{b}_{\rm e} \dfrac{\hat{\delta}W_{\rm e}}{\hat{\delta}\bm{b}_{\rm e}} \!\!$ \\[1.5ex]
      	$\bm{\tau}' = \bm{\tau}-\frac{1}{3} \left( \tr{\bm{\tau}} \right) \bm{I}$ \\
      	$\phi=\sqrt{\frac{3}{2}\bm{\tau}' \cdot \bm{\tau}'} - \sigma_{\rm y} (\gamma)$ \rule{0em}{3.1ex} \\
      	$\bm{n}=\dfrac{\partial \phi}{\partial \bm{\tau}}$  \hspace*{7.0cm} $\triangleright\;\text{automation: } \bm{n} \leftarrow \dfrac{\hat{\delta}\phi}{\hat{\delta}\bfs{\tau}} \!\!$ \\[1.5ex]
      	$\bm{\mathcal{Z}} = \bm{F}\bm{C}_{{\rm p}}^{-1} - \exp{\left( -2 \left( \gamma-\gamma_n \right)\bm{n} \right)}\bm{F}\bm{C}_{{\rm p},n}^{-1}$ \rule[-1.5ex]{0em}{4.2ex} \\ \hline
      	$\brm{h}=\left\{ {C}_{{\rm p},11}^{-1} - 1 , {C}_{{\rm p},22}^{-1} - 1 , {C}_{{\rm p},33}^{-1} - 1 , {C}_{{\rm p},23}^{-1} , {C}_{{\rm p},13}^{-1} , {C}_{{\rm p},12}^{-1} , \gamma \right\}$ \rule{0em}{3ex} \\
      	$\brm{Q}=\left\{ \mathcal{Z}_{11} , \mathcal{Z}_{22} , \mathcal{Z}_{33} , \mathcal{Z}_{23} , \mathcal{Z}_{13} , \mathcal{Z}_{12} , \phi \right\}$ \rule[-1.2ex]{0em}{4ex} \\ \hline
	\end{tabular}
	}
\end{mybox}

Consider now a simple laminate in which both phases are governed by an elastic-plastic material model. Expressing the local deformation gradients in terms of $\bar{\bm{F}}$ and $\bm{c}$, as in Eq.~\eqref{eq:Fi}, the macroscopic elastic strain energy $\bar{W}=\langle W\rangle$ reads
\begin{equation}
\bar{W}(\bar{\bm{F}},\bm{c},\brm{h}_1,\brm{h}_2)={(1-\eta) W_1(\bm{F}_1,\brm{h}_1)+\eta W_2(\bm{F}_2,\brm{h}_2)} ,
\end{equation}
and the macroscopic stress is obtained as
\begin{equation}
\bar{\bm{P}}
  =\frac{\partial\bar{W}(\bar{\bm{F}},\bm{c},\brm{h}_1,\brm{h}_2)}{\partial\bar{\bm{F}}} 
  ={(1-\eta)\frac{\partial W_1}{\partial\bm{F}_1}\frac{\partial\bm{F}_1}{\partial\bar{\bm{F}}}
    +\eta\frac{\partial W_2}{\partial\bm{F}_2}\frac{\partial\bm{F}_2}{\partial\bar{\bm{F}}}
  =(1-\eta)\bm{P}_1+\eta\bm{P}_2} .
\end{equation}

The unknown vector $\bm{c}$ is obtained by solving, using the Newton method, the compatibility condition~\eqref{eq:comp:P} written here in the residual form as
\begin{equation}
\brm{R}(\bar{\bm{F}},\bm{c},\brm{h}_1,\brm{h}_2)=(\bm{P}_2-\bm{P}_1)\bm{N}=\bm{0} ,
\end{equation}
where the internal variables $\brm{h}_i$ depend on $\bm{c}$ through $\bm{F}_i$, thus $\brm{h}_i=\brm{h}_i(\bm{F}_i(\bar{\bm{F}},\bm{c}))$, and this dependence must be taken into account when the residual $\brm{R}$ is linearized.
The complete computational scheme is summarized in Algorithm~\ref{alg:2}, which includes consistent linearization of the nested iterative-subiterative scheme. In particular, once the implicit dependencies are correctly identified and introduced into the code, the consistent overall tangent (denoted as $\bar{\mathbb{L}}^{\rm alg}$ in Algorithm~\ref{alg:2}) is obtained as the computational derivative $\hat{\delta}\bar{\bm{P}}/\hat{\delta}\bar{\bm{F}}$.



\begin{algorithm}
\caption{\newline AD-based formulation of the incremental scheme for
an elasto-plastic two-phase composite} 
\label{alg:2}
\begin{spacing}{1.2}
\begin{algorithmic}
  \State {\bf input:} $\bar{\bm{F}}$, $\bm{c}^n$, $\brm{h}_1^n$,
    $\brm{h}_2^n$ \rule{0em}{3ex}
  \State $\bm{c} \gets \bm{c}^n$
  \Repeat
  \State $\bm{F}_1 \gets \bar{\bm{F}} - \eta \bm{c} \otimes \bm{N}$
  \State $\bm{F}_2 \gets \bar{\bm{F}} + (1-\eta) \bm{c} \otimes \bm{N}$
  \State $\{\brm{h}_1, \bm{P}_1, \brm{G}_1 \} \gets \mathtt{StateUpdate}\left[ \bm{F}_1, \brm{h}_1^n \right]$
  \State $\{\brm{h}_2, \bm{P}_2, \brm{G}_2 \} \gets \mathtt{StateUpdate}\left[ \bm{F}_2, \brm{h}_2^n \right]$
  \State $\brm{R} \gets \left( \bm{P}_2 - \bm{P}_1 \right) \bm{N}$
  \State $\brm{B} \gets \dfrac{\hat{\delta} \brm{R}}{\hat{\delta} \bm{c}}$
    \Comment{tangent matrix, $\brm{B}=\dfrac{\partial\brm{R}}{\partial\bm{c}}$}
    \vspace{0.5ex}
  \State $\Delta \bm{c} \gets -\brm{B}^{-1} \brm{R}$
  \State $\bm{c} \gets \bm{c} + \Delta \bm{c}$
  \Until $\| \Delta \bm{c} \| \le tol$ \vspace{0.5ex}
  \State $\bm{c} \gets \bm{c}
    \Big|_{\gfrac{\RD \bm{c}}{\RD \bar{\bm{F}}}
    =-\brm{B}^{-1} \gfrac{\hat{\delta}\brm{R}}{\hat{\delta}\bar{\bm{F}}}
    \big|_{\bm{c}={{\rm const}}}}$ \vspace{0.5ex}
  \State $\bm{F}_1 \gets \bar{\bm{F}} - \eta \bm{c} \otimes \bm{N}$
  \State $\bm{F}_2 \gets \bar{\bm{F}} + (1-\eta) \bm{c} \otimes \bm{N}$
  \State $\brm{h}_1 \gets \brm{h}_1 \Big|_{\gfrac{\RD \brm{h}_1}{\RD\bm{F}_1} = \brm{G}_1}$
    \Comment{introduce the implicit dependence of $\brm{h}_1$ on $\bm{F}_1$}
  \State $\brm{h}_2 \gets \brm{h}_2 \Big|_{\gfrac{\RD \brm{h}_2}{\RD\bm{F}_2} = \brm{G}_2}$ 
    \Comment{introduce the implicit dependence of $\brm{h}_2$ on $\bm{F}_2$}\vspace{0.5ex}
  \State $\bar{W} \gets \left( 1-\eta \right) W_1(\bm{F}_1,\brm{h}_1) + \eta W_2(\bm{F}_2,\brm{h}_2)$
    \vspace{0.5ex}
  \State $\bar{\bm{P}} \gets
  	 \left.\dfrac{\hat{\delta}\bar{W}}{\hat{\delta}\bar{\bm{F}}} \right|_{
  	 \bm{c}=\rm{const} , \, \brm{h}_1={\rm const} , \, \brm{h}_2={\rm const}}$
    \Comment{AD exception ensures that $\bar{\bm{P}}$ is computed correctly}
  \State $\bar{\mathbb{L}}^{\rm alg} \gets \dfrac{\hat{\delta}\bar{\bm{P}}}{\hat{\delta}\bar{\bm{F}}}$
        \vspace{0.5ex}
  \State {\bf return:} $\bm{c}$, $\bar{\bm{P}}$, $\bar{\mathbb{L}}^{\rm alg}$, $\brm{h}_1$, $\brm{h}_2$
\end{algorithmic}
\end{spacing}
\end{algorithm}

In practice, the iterative Newton scheme in Algorithm~\ref{alg:2} can be enhanced by a line search technique which improves the robustness of the computational scheme. The related details are omitted here.